\documentclass[journal]{IEEEtran}

\ifCLASSINFOpdf
   \usepackage[pdftex]{graphicx}
\else
\fi

\usepackage{comment}

\usepackage{amsmath}
\usepackage{mathtools}

\usepackage{acronym}

\usepackage{amsfonts}
\usepackage{amsthm}
\usepackage{thmtools}

\usepackage{siunitx}

\usepackage{mleftright}
\usepackage{bm}
\usepackage{bbm}

\usepackage{placeins}
\usepackage{tikz}
\usepackage{pgfplots}

\usepackage[textsize=tiny, disable]{todonotes} %

\usetikzlibrary{external}
\definecolor{tab101}{RGB}{228,26,28}
\definecolor{tab102}{RGB}{55,126,184}
\definecolor{tab103}{RGB}{77,175,74}
\definecolor{tab104}{RGB}{152,78,163}
\definecolor{tab105}{RGB}{255,127,0}

\definecolor{cR1}{rgb}{0,.59,.51}  %
\definecolor{cR2}{rgb}{.27,.39,.66} %
\definecolor{cR3}{rgb}{.63,.13,.13} %

\definecolor{cR4}{rgb}{.64,.06,.48} %
\definecolor{cR5}{rgb}{.87,.60,.10} %
\definecolor{cR6}{rgb}{.14,.63,.87} %

\definecolor{cR7}{RGB}{100, 221, 23} %
\definecolor{cR8}{RGB}{167,130,46} %
\definecolor{cR9}{rgb}{0.4, 0.6, 0.2} %

\ifCLASSOPTIONcompsoc
  \usepackage[caption=false,font=normalsize,labelfont=sf,textfont=sf]{subfig}
\else
  \usepackage[caption=false,font=footnotesize]{subfig}
\fi

\usepackage{graphicx}
\usepackage{color}
\usepackage{booktabs}
\usepackage{pgf, tikz, pgfplots}
\usetikzlibrary{shapes,arrows.meta, calc}

\usepackage[ruled,vlined, linesnumbered]{algorithm2e}
\let\oldnl\nl%
\newcommand{\nonl}{\renewcommand{\nl}{\let\nl\oldnl}}%

\newcommand\vek[1]{{\bm{#1}}}
\newcommand\rv[1]{{\textsf{#1}}}
\newcommand\rvvek[1]{{\textsf{\textbf{#1}}}}
\newcommand\ebno{E_{\text{b}}/N_0}

\DeclareMathOperator*{\argmax}{\arg\!\max}

\DeclareMathOperator*{\maxstar}{max^\star}

\definecolor{KITgreen}{rgb}{0,.59,.51}
\definecolor{KITpalegreen}{RGB}{130,190,60} 
\definecolor{KITblack}{rgb}{0,0,0}
\definecolor{KITblue}{rgb}{.27,.39,.66}
\definecolor{KITred}{rgb}{.63,.13,.13}
\definecolor{KITpurple}{rgb}{.64,.06,.48}
\definecolor{KITcyan}{rgb}{.14,.63,.87}
\definecolor{KITyellow}{rgb}{.98,.89,0}
\definecolor{KITorange}{rgb}{.87,.60,.10}

    \acrodef{AIR}[AIR]{achievable information rate}
    \acrodef{APP}[APP]{a posteriori probability}
    \acrodef{AWGN}[AWGN]{additive white Gaussian noise}
    \acrodef{BER}[BER]{bit error rate}
    \acrodef{BICM}[BICM]{bit-interleaved coded modulation}
    \acrodef{BMD}[BMD]{bit-metric decoder}
    \acrodef{BMI}[BMI]{bitwise mutual information}
    \acrodef{BP}[BP]{belief propagation}
    \acrodef{BPSK}[BPSK]{binary phase-shift keying}
    \acrodef{BSC}[BSC]{binary symmetric channel}
    \acrodef{CEL}[CEL]{Communications Engineering Lab}
    \acrodef{CM}[CM]{Coded Modulation}
    \acrodef{CPU}[CPU]{central processing unit}
    \acrodef{DFE}[DFE]{decision-feedback equalization}
    \acrodef{DL}[DL]{deep learning}
    \acrodef{DNN}[DNN]{deep neural network}
    \acrodef{FN}[FN]{factor node}
    \acrodef{FFG}[FFG]{Forney-style factor graph}
    \acrodef{FEC}[FEC]{forward error correction}
    \acrodef{FIR}[FIR]{finite impulse response}
    \acrodef{GMI}[GMI]{generalized mutual information}
    \acrodef{GPU}[GPU]{graphics processing unit}
    \acrodef{GNN}[GNN]{graph neural network}
    \acrodef{iid}[i.i.d.]{independent and identically distributed}
    \acrodef{ISI}[ISI]{inter-symbol interference}
    \acrodef{KIT}[KIT]{Karlsruhe Institute of Technology}
    \acrodef{LDPC}[LDPC]{low-density parity-check}
    \acrodef{LLR}[LLR]{log-likelihood ratio}
    \acrodef{LR}[LR]{likelihood ratio}
    \acrodef{MAP}[MAP]{maximum a posteriori}
    \acrodef{MIMO}[MIMO]{multiple-input multiple-output}
    \acrodef{ML}[ML]{maximum likelihood}
    \acrodef{MLSE}[MLSE]{maximum likelihood sequence estimation}
    \acrodef{MMSE}[LMMSE]{linear minimum mean squared error}
    \acrodef{MSE}[MSE]{mean squared error}
    \acrodef{NBP}[NBP]{neural belief propagation}
    \acrodef{NN}[NN]{neural network}
    \acrodef{pdf}[PDF]{probability density function}
    \acrodef{pmf}[PMF]{probability mass function}
    \acrodef{QAM}[QAM]{quadrature amplitude modulation}
    \acrodef{RV}[RV]{random variable}
    \acrodef{SER}[SER]{symbol error rate}
    \acrodef{SISO}[SISO]{soft-input soft-output}
    \acrodef{SNR}[SNR]{signal-to-noise ratio}
    \acrodef{SPA}[SPA]{sum-product algorithm}
    \acrodef{SPC}[SPC]{single parity check}
    \acrodef{TEQ}[TEQ]{turbo-equalization}
    \acrodef{VN}[VN]{variable node}

\pgfplotsset{compat=newest}

\begin{document}
\title{Low-complexity Near-optimum Symbol Detection Based on Neural Enhancement of Factor Graphs}

\author{Luca Schmid and Laurent Schmalen,~\IEEEmembership{Senior~Member,~IEEE}%
\thanks{This work has received funding in part from the European Research Council (ERC) under the European Union’s Horizon 2020 research and innovation programme (grant agreement No. 101001899) and in part from the German
Federal Ministry of Education and Research (BMBF) within the project Open6GHub (grant agreement 16KISK010).
Parts of this paper have been presented at the IEEE International Workshop on Signal Processing Advances in Wireless Communications (SPAWC), Oulu, Finland, July 2022 in paper~\cite{schmid_neural_2022}.}
\thanks{Both authors are with Karlsruhe Institute of Technology (KIT), Communications Engineering Lab (CEL), Hertzstr. 16, 76187 Karlsruhe, Germany. E-mail: \texttt{luca.schmid@kit.edu}, \texttt{schmalen@kit.edu}}%
}

\markboth{Submitted version, \today}%
{Submitted version, \today}

\maketitle

\vspace{-1.5em}
\begin{abstract}
We consider the application of the factor graph framework for symbol detection on 
linear inter-symbol interference channels.
Based on the Ungerboeck observation model, 
a detection algorithm with appealing complexity properties can be derived.
However, since the underlying factor graph contains cycles, the \ac{SPA} yields a suboptimal algorithm.
In this paper, we develop and evaluate efficient strategies to improve the performance of
the factor graph-based symbol detection by means of neural enhancement.
In particular, we consider neural belief propagation and generalizations of the factor nodes as an effective way to mitigate the effect of
cycles within the factor graph.
By applying a generic preprocessor to the channel output, 
we propose a simple technique to vary the underlying factor graph in every \ac{SPA} iteration. 
Using this dynamic factor graph transition, we intend to preserve the
extrinsic nature of the \ac{SPA} messages which is otherwise impaired due to cycles.
Simulation results show that the proposed methods can
massively improve the detection performance, 
even approaching the maximum a posteriori performance for various transmission scenarios, 
while preserving a complexity which is linear in both the block
length and the channel memory.
\end{abstract}

\begin{IEEEkeywords}
    \noindent Factor graphs, neural belief propagation, symbol detection, channels with memory,
    high-level parallelism
\end{IEEEkeywords}
\acresetall

\IEEEpeerreviewmaketitle

\section{Introduction}
The well-known task of data transmission over a channel with
linear \ac{ISI} impaired by \ac{AWGN} is considered in this paper.
\ac{ISI} is ubiquitous in many wireline and wireless communication systems where, e.g., multipath propagation is caused by reflections and refraction of the transmit signal in the wireless channel. Left uncompensated, \ac{ISI} leads to a distortion of the signal and causes high error rates at the receiver~\cite[Chap.~9]{proakis_digital_2007}.
Detection algorithms are thus required at the receiver in order to recover the original transmit signal.
Optimum detection with respect to the symbol error probability is based
on \ac{MAP} symbol detection. The BCJR algorithm~\cite{bahl_optimal_1974}
is an efficient \ac{MAP} algorithm whose complexity is linear in the block length but
exponential in the memory of the channel and the number of bits mapped to each constellation symbol.
In many practical scenarios where channels have
large memory or where high-order constellations are used,
the BCJR algorithm becomes prohibitively complex.
Therefore, the development of computationally efficient algorithms with near-optimum performance has become a major field of research.
Classical low-complexity equalizers like linear transversal filters and 
algorithms based on \ac{DFE} yield acceptable performance for a wide range of communication channels
with well behaved spectral characteristics, but perform poorly for channels with severe \ac{ISI}
and spectral zeros~\cite[Sec.~9.4]{proakis_digital_2007}.
A common approach for reduced-complexity \ac{MAP} symbol detection 
is a simplification of the trellis search within the 
BCJR algorithm~\cite{fertonani_reduced-complexity_2007}.
Either a reduced search on the full-complexity trellis can be
performed (e.g., the $M$-BCJR algorithm~\cite{franz_concatenated_1998}), or
the number of trellis states can be reduced
(e.g., the RS-BCJR algorithm~\cite{colavolpe_reduced-state_2001}).
However, these algorithms reduce the complexity of the BCJR algorithm only by
a scalar factor and the performance-complexity tradeoff is only satisfactory for a particular subset of \ac{ISI} channels~\cite{fertonani_reduced-complexity_2007}.
An alternative approach to reduce the detection complexity is channel shortening~\cite{rusek_optimal_2012}. Filtering the channel output with a channel shortening filter and then applying the BCJR algorithm on the shortened channel model enables a significantly reduced detection complexity but potentially comes with a performance-complexity tradeoff.

The advent of suboptimal iterative decoding in the context of turbo codes and LDPC codes has led to a rediscovery of message passing on graphical models. 
The powerful \emph{factor graph} framework~\cite{kschischang_factor_2001} provides a universal modeling tool for algorithms with controllable complexity.
Based on the Ungerboeck observation model~\cite{ungerboeck_adaptive_1974},
Colavolpe et al.\ employed factor graphs and the \ac{SPA} to derive a symbol detector with substantially 
reduced complexity~\cite{colavolpe_siso_2011}.
In particular, the complexity is linear both in the block length
and the channel memory.
The proposed algorithm is however suboptimal, since its underlying factor graph contains cycles.

Recently, model-based deep learning has shown great potential to empower various suboptimal communication algorithms~\cite{shlezinger_model-based_2020} and overcome their limitations.
In~\cite{shlezinger_data-driven_2020}, the factor nodes of a cycle-free factor graph are replaced by \acp{DNN} that are utilized to learn the local mappings of the factor nodes, thereby robustifying the algorithm towards model uncertainties. However, the algorithm still suffers from a complexity which is comparable to the complexity of the BCJR algorithm. Therefore, we focus on model-based deep learning approaches which are based on the Ungerboeck observation model, in the following. 
To mitigate the performance loss for cyclic factor graphs,
a \ac{GNN}, which is structurally identical to the original graph but has fully parametrized message updates, is proposed in~\cite{satorras_neural_2021}.
The \ac{GNN} runs conjointly to the original algorithm
and corrects the \ac{SPA} messages after each iteration.
The authors in~\cite{liu_novel_2021} compensate the performance degradation due to cycles in the graph by concatenating a supplemental \ac{NN}-based \ac{FN} to the factor graph. This additional \ac{FN} is connected to all \acp{VN} and is optimized in an end-to-end manner. However, the underlying \ac{NN} structure is specifically tailored to binary transmission which substantially limits its scope of application.
Instead of replacing different components of the factor graph by \acp{DNN}, the \ac{SPA} is unfolded to a \ac{DNN} and the resulting graph is equipped with tunable weights in~\cite{nachmani_learning_2016}. This approach is known as \ac{NBP}.

This paper aims at closing the gap between optimum and low-complexity symbol detection.
We consider \ac{NBP} on the Ungerboeck-based factor graph and further generalize the graph by introducing additional multiplicative weights within the \acp{FN}.
Optimizing all weights in an end-to-end manner leads to considerable performance gains.
Moreover, we leverage the high sensibility of the \ac{SPA} to a variation of the underlying graph by applying an optimizable linear filter to the channel output, which allows us to modify the observation model and thereby the factor graph itself.
By a combination of multiple factor graph instances in parallel, as well a dynamic variation of the graphs over the message passing iterations, we exploit this graph diversity in order to significantly improve the overall detection performance, and close the gap to optimum symbol detection for a variety of channels.\\

The remainder of this paper is organized as follows.
In Section~\ref{sec:factor_graph}, we briefly review the concept of factor graphs and the \ac{SPA}~\cite{kschischang_factor_2001}. 
Section~\ref{sec:channel_model} formulates the fundamental problem of symbol detection 
on channels with linear \ac{ISI}.
Using the factor graph framework, we present a suboptimal, but low-complexity, symbol detection algorithm. 
This provides the basis for Section~\ref{sec:neural_enhancement}, 
in which we propose and discuss various generalizations and enhancements to the algorithm.
Section~\ref{sec:results} examines the behavior of the proposed algorithms
for different linear \ac{ISI} channels and quantifies its performance compared to existing symbol detectors.
Some concluding remarks are given in Section~\ref{sec:conclusion}.

\subsection*{Notation}
Throughout the paper, we use bold letters for
non-scalar quantities. 
Upper case letters denote matrices $\vek{X}$ 
and $X_{m,n}$ represents the entry at row $m$ and column $n$.
Lower case letters are used for 
column vectors
$\vek{x}$. The $i$th element of $\vek{x}$ is written as $x_i$ and 
the stacking of a vector from multiple scalars is denoted by
$[ x_i ]_{i=1}^{n} = \vek{x}$.
$\lVert \cdot \rVert$ denotes the Euclidean norm and $(\cdot)^{\textrm{H}}$ is the
conjugate transpose (Hermitian) operator.
The computation of the term $\ln\mleft(\mathrm{e}^{\delta_1} + \ldots + \mathrm{e}^{\delta_n} \mright)$
can be carried out using the Jacobian logarithm~\cite{robertson_comparison_1995} and is denoted
by $\maxstar\limits_{i} \delta_i$.
The probability measure of a \ac{RV} $\rv{x}$ evaluated at $x$ is denoted by ${P_{\rv{x}}(\rv{x} = x)}$. If it is clear from the context, we may use the shorter notation
${P(\rv{x}=x)}$ for the sake of simplicity.
The \ac{pdf} of a continuous \ac{RV} $\rv{y}$ is denoted by $p_{\rv{y}}(y)$ or $p(y)$.
The \ac{pmf} of a discrete \ac{RV} $x$ is $P_{\rv{x}}(x)$ or $P(x)$.
The Gaussian distribution, characterized by its mean $\mu$ and variance $\sigma^2$, 
is written as $\mathcal{N}(\mu,\sigma^2)$.
The expected value of an \ac{RV} $\rv{x}$ is denoted by 
$\mathbb{E}_{\rv{x}} \mleft \{ \rv{x} \mright \}$ and the mutual information between the \acp{RV} $\rv{x}$ and $\rv{y}$ is $I(\rv{x};\rv{y})$.
We use calligraphic letters to denote a set $\mathcal{X}$ of cardinality $|\mathcal{X}|$.

\section{Factor Graphs and Marginalization}\label{sec:factor_graph}
The framework of factor graphs and the \ac{SPA} is a flexible tool for
algorithmic modeling of efficient inference algorithms.
By representing the factorization of a composite global function of many variables
in a graphical way, the computation of various marginalizations of this function
can be efficiently implemented by a message passing algorithm.
Since factor graphs are the foundation of our work, we review the basic concepts in this section.

Let $f(\mathcal{X})$ be a so-called ``global'' function which depends on a set of variables ${\mathcal{X} = \left\{ x_0, x_1, \ldots, x_n \right\}}$ with $x_i \in \mathcal{M}$ for $i=0,\ldots,n$.
The marginalization of $f(\mathcal{X})$ towards a single marginal variable $x_i$ 
typically requires $|\mathcal{M}|^n$ additions which
can quickly become computationally infeasable for large $n$.
The complexity can be significantly reduced by means of the distributive law
if the global function is factorizable~\cite{kschischang_factor_2001}.
Let us assume that $f(\mathcal{X})$ can be factorized as
\begin{equation} \label{eq:factorization}
    f(\mathcal{X}) = \prod\limits_{j=1}^J f_j(\mathcal{X}_j),\quad \mathcal{X}_j \subset \mathcal{X},
\end{equation}
where each factor $f_j(\mathcal{X}_j)$ only depends on a subset of the variables $\mathcal{X}_j$.
A \emph{factor graph} represents the factorization of a multivariate function
in a graphical way~\cite{loeliger_introduction_2004}. 
The following rules define the bijective relationship between the generic factorization 
in \eqref{eq:factorization} and its corresponding factor graph:
\begin{itemize}
    \item Every factor $f_j(\mathcal{X}_j)$ is represented by a unique vertex, the so-called \ac{FN} $f_j$.
    \item Every variable $x \in \mathcal{X}$ is represented by a unique vertex, the so-called \ac{VN} $x$.
    \item An \ac{FN} $f_j$ is connected to a \ac{VN} $x$
          if and only if the corresponding factor $f_j(\mathcal{X}_j)$ is a function of $x$, 
          i.e., if $x \in \mathcal{X}_j =: \mathcal{N}(f_j)$.
\end{itemize}
The notation $\mathcal{N}(f_j)$ denotes the \emph{neighborhood} of the \ac{FN} and is introduced to emphasize the fact that all variables
on which a factor depends are represented by adjacent \acp{VN} in the factor graph.
Equivalently, ${\mathcal{N}(x) := \{ f_j , \, j=1,\ldots,J: x \in \mathcal{N}(f_j) \}}$ denotes the neighborhood of the \ac{VN} $x$.
It is worth mentioning that the factor graph representation of a factorization is unique with respect to the structure of the resulting graph. However, there can be various factorizations of the same global function, leading to disparate factor graph representations~\cite{kschischang_factor_2001}.

The \ac{SPA} is a message passing algorithm which computes
the marginalization $f(x_i)$ of the global function $f(\mathcal{X})$ towards each variable
${x_i \in \mathcal{X}}$, respectively. It implicitly
leverages the distributive law on the factorization of $f(\mathcal{X})$.
Messages are propagated between the nodes of the factor graph along its edges 
and represent interim results of the marginalization. 
Let $\mu_{f_j \rightarrow x}(x)$ denote a message sent 
from \ac{FN} $f_j$ along an outgoing edge to \ac{VN} $x$.
Consequently, $\mu_{x \rightarrow f_j}(x)$ denotes  a message on the same edge, but sent in the
opposite direction.
The message passing algorithm is based on one central message update rule for the \acp{VN} and
\acp{FN}, respectively.
In the logarithmic domain\footnote{When it comes to hardware implementation,
it can be advantageous to carry out the SPA in the logarithmic domain due to less numerical instabilities and a reduced computational complexity.}, 
the message updates are
\begin{align}
\mu_{x \rightarrow f_j}(x)
    &= \sum\limits_{f' \in \mathcal{N}(x) \setminus \{ f_j \}} \mu_{f' \rightarrow x} (x) \label{eq:v2f_update} \\
    \mu_{f_j \rightarrow x}(x)
    &= \maxstar\limits_{\sim \{ x \}}
    \mleft( \ln \mleft( f_j(\mathcal{X}_j) \mright) + \sum\limits_{\mathclap{x' \in \mathcal{N}(f_j) \setminus \{ x \}}} \mu_{x' \rightarrow f_j} (x') \mright). \label{eq:f2v_update}
\end{align}
They define the computation of an outgoing message,
given the incoming messages on all \emph{other} incident edges of a node~\cite{kschischang_factor_2001}, called \emph{extrinsic messages}. 
The local marginalization in the \ac{FN} to \ac{VN} update, i.e., the Jacobian algorithm over all extrinsic variables, is thereby denoted by the
\emph{summary} operator $\maxstar_{\sim \{ x \}}$.

Based on the \ac{SPA} message update rule, we can compute marginals by propagating messages through the respective factor graph.
If the graph is tree-structured, messages travel forward and backward through the entire graph, starting at the leaf nodes.
Based on the computed messages, the exact marginals
\begin{equation*}
    f(x_i) = \exp \mleft( \sum\limits_{f' \in \mathcal{N}(x_i)} \mu_{f' \rightarrow x_i}(x_i) \mright)
\end{equation*}
can be obtained.
Since the message updates are local~\cite{kschischang_factor_2001}
and because the \ac{SPA} makes no reference to the topology of the graph~\cite{yedidia_understanding_2003},
the \ac{SPA} may also be applied to factor graphs with cycles, yielding an iterative algorithm.
The messages are initialized with an unbiased state in iteration
$n=0$ and are iteratively updated by following a certain schedule
until convergence or a stopping criterion is reached.
In the case of cyclic factor graphs, the superscript ${}^{(n)}$, with ${n=0,\ldots,N}$, indicates the iteration in which the message
$\mu_{a \rightarrow b}^{(n)}$ is computed.
In general, convergence of the \ac{SPA} on cyclic factor graphs is not guaranteed and the iterative algorithm only yields an approximation 
\begin{equation}\label{eq:loopy_marginal}
    \hat{f}(x_i) := \exp \mleft( \sum\limits_{f' \in \mathcal{N}(x_i)} \mu_{f' \rightarrow x_i}^{(N)}(x_i) \mright)
\end{equation}
of the exact marginal $f(x_i)$~\cite{kschischang_factor_2001}.
However, many successful applications, e.g., decoders of error-correcting codes, are
based on message passing algorithms on cyclic graphs.

\section{Symbol Detection}\label{sec:channel_model}
We consider the transmission of an information sequence 
${\rvvek{c} = [\rv{c}_k]_{k=1}^{K} \in \mathcal{M}^K}$ of a multilevel
constellation ${\mathcal{M}} = {\{\text{m}_i \in \mathbb{C}, i=1,\ldots,M \}}$ 
over a complex baseband channel, impaired by linear interference and \ac{AWGN}.
The bit pattern of length ${m := \log_2 \mleft( M \mright)}$ which corresponds to a 
symbol $\rv{c}_k$ 
is denoted by ${\rvvek{b}(\rv{c}_k) = [\rv{b}_{i}(\rv{c}_k) ]_{i=1}^{m}}$.
The relationship between the \ac{iid} 
information symbols $\rv{c}_k$ and the receive symbols $\rv{y}_k$ can be expressed by an equivalent discrete-time channel model~\cite{forney_lower_1972}:
\begin{equation}\label{eq:convolution}
    \rv{y}_k = \sum\limits_{\ell=0}^{L} h_{\ell} \rv{c}_{k-\ell} + \rv{w}_k.
    \quad k = 1,\ldots,K+L.
\end{equation}
For a channel with memory $L$, $\vek{h} \in \mathbb{C}^{L+1}$ is the finite channel impulse response  
and $\rvvek{w} \sim \mathcal{C}\mathcal{N}(0,\sigma^2 \boldsymbol{I})$ 
denotes white circular Gaussian noise.
The channel is assumed to be static, which leads to the channel impulse response $\vek{h}$ 
being constant over time.
The symbols $\rv{c}_k$ for ${k<1}$ and ${k > K}$ are information symbols from the same constellation $\mathcal{M}$. 
We assume that these boundary symbols are fully known at the receiver, as they are either pilot symbols or information symbols from an adjacent transmission block which was already detected and decoded.
An equivalent transmit sequence is given by
$\check{\rvvek{c}} := [\rv{c}_k]_{k=1-L}^{K+L} \in \mathcal{M}^{(K+2L)}$.
Since the interference is linear, \eqref{eq:convolution} can
be described in matrix vector notation:
\begin{equation*}
    \rvvek{y} = \vek{H} \check{\rvvek{c}} + \rvvek{w}.
\end{equation*}
The matrix ${\vek{H} \in \mathbb{C}^{(K+L) \times (K+2L)}}$ is a band-structured Toeplitz matrix which represents the convolution of the transmit sequence $\check{\rvvek{c}}$ with the channel impulse response $\vek{h}$.

We study the problem of symbol detection, i.e., the estimation of the information symbols $\rv{c}_k$, ${k=1,\ldots,K}$ from a sequence $\rvvek{y}$, observed at the receiver. 
In the context of Bayesian inference, 
we are interested in the
\ac{APP} distribution 
\begin{equation*}
    P(\vek{c|y}) \propto p(\vek{y|c})P(\vek{c}),
\end{equation*}
where proportionality $\propto$ denotes two terms only differing in a factor independent of $\rvvek{c}$.
The \ac{APP} can be factored into the likelihood 
\begin{equation}\label{eq:likelihood}
p(\vek{y}|\vek{c}) = \frac{1}{(\pi \sigma^2)^K} \exp \left ( -\frac{\Vert \vek{y-Hc} \Vert ^2}{\sigma^2} \right )
\end{equation}
and the a priori probability  
$P(\vek{c})$, using Bayes' theorem~\cite[Chap.~2]{harney_bayesian_2003}.
The symbol-wise \acp{APP} are obtained by computing the marginals
\begin{equation}\label{eq:symbol_wise_app}
    P(\rv{c}_k = c |\vek{y}) = 
    \sum_{\substack{\vek{c} \in \mathcal{M}^K \\ c_k = c}} P(\rvvek{c}=\vek{c|y}), \quad k=1,\ldots,K,
\end{equation}
on which symbol detection can be based.
In case of hard decision, the symbol-wise \ac{MAP} detection 
\begin{equation*}
    \hat{\rv{c}}_k =  \argmax_{c \in \mathcal{M}} P(\rv{c}_k = c |\vek{y}), \quad k=1,\ldots,K
\end{equation*}
yields the minimum probability of error for each symbol
decision, respectively~\cite[Sec.~9.3]{proakis_digital_2007}.

\subsection{Factor Graph Modeling}\label{sec:ubfgeq}
\begin{figure}[tb]
    \centering
    \tikzstyle{fn} = [draw, very thick, regular polygon, regular polygon sides=4, minimum width = 3em, inner sep=0pt]
\tikzstyle{fn_ghost} = [very thick, regular polygon, regular polygon sides=4, minimum width = 3em, inner sep=0pt]
\tikzstyle{vn} = [draw, very thick, circle, inner sep=0pt, minimum size = 2em]
\tikzstyle{vn_ghost} = [very thick, circle, inner sep=0pt, minimum size = 2em]

\begin{tikzpicture}[auto, node distance=3em and 3.3em, thick]
\clip (1.7, -2.2) rectangle (9.7, 2.0);
    \node [vn_ghost] (c0){};
    \node [vn_ghost, right= of c0] (c1){$\cdots$};
    \node [vn, label=center:$c_{2}$, right= of c1] (c2){};
    \node [vn, label=center:$c_{3}$, right= of c2] (c3){};
    \node [vn, label=center:$c_{4}$, right= of c3] (c4){};
    \node [vn_ghost, right= of c4] (c5){$\cdots$};
    \node [vn_ghost, right= of c5] (c6){};

    \node [fill=KITgreen, fill opacity=0.2, fn, label=center: $P_{2}$, above=2em of c2.100, anchor = south east] (p2){};
    \node [fill=KITgreen, fill opacity=0.2, fn, label=center: $P_{3}$, above=2em of c3.100, anchor = south east] (p3){};
    \node [fill=KITgreen, fill opacity=0.2, fn, label=center: $P_{4}$, above=2em of c4.100, anchor = south east] (p4){};
    \node [fill=KITblue, fill opacity=0.2, fn, label=center: $F_{2}$, above=2em of c2.80, anchor = south west] (f2){};
    \node [fill=KITblue, fill opacity=0.2, fn, label=center: $F_{3}$, above=2em of c3.80, anchor = south west] (f3){};
    \node [fill=KITblue, fill opacity=0.2, fn, label=center: $F_{4}$, above=2em of c4.80, anchor = south west] (f4){};
    \node [fill=KITred, fill opacity=0.2, fn, label=center: $I_{2,0}$, below= of c2.260, anchor = north east] (I20){};
    \node [fill=KITred, fill opacity=0.2, fn, label=center: $I_{2,1}$, below= of c2.280, anchor = north west] (I21){};
    \node [fill=KITred, fill opacity=0.2, fn, label=center: $I_{3,1}$, below= of c3.260, anchor = north east] (I31){};
    \node [fill=KITred, fill opacity=0.2, fn, label=center: $I_{3,2}$, below= of c3.280, anchor = north west] (I32){};
    \node [fill=KITred, fill opacity=0.2, fn, label=center: $I_{4,2}$, below= of c4.260, anchor = north east] (I42){};
    \node [fill=KITred, fill opacity=0.2, fn, label=center: $I_{4,3}$, below= of c4.280, anchor = north west] (I43){};
    \node [fn_ghost, below= of c5.260, anchor = north east] (I53){};
    \node [fn_ghost, below= of c5.280, anchor = north west] (I54){};
    \node [fn_ghost, below= of c6.260, anchor = north east] (I64){};
    \node [fn_ghost, below= of c6.280, anchor = north west] (I65){};
    
    \draw[-] (c2.north) -- (p2.south);
    \draw[-] (c2.north) -- (f2.south);
    \draw[-] (c3.north) -- (p3.south);
    \draw[-] (c3.north) -- (f3.south);
    \draw[-] (c4.north) -- (p4.south);
    \draw[-] (c4.north) -- (f4.south);
    \draw[-] (I20.north) -- (c2.south);
    \draw[dotted] (I20.north) -- ($(I20.north)!0.4!(c0.south)$);
    \draw[-] (I20.north) -- ($(I20.north)!0.2!(c0.south)$);
    \draw[-] (I21.north) -- (c2.south);
    \draw[-] (I21.north) -- ($(I21.north)!0.58!(c1.south)$);
    \draw[dotted] (I21.north) -- ($(I21.north)!1.0!(c1.south)$);
   \draw[-] (c2.south) -- (I32.north);
    \draw[-] (c2.south) -- (I42.north);
    \draw[-] (I31.north) -- (c3.south);
    \draw[-] (I31.north) -- ($(I31.north)!0.7!(c1.south)$);
    \draw[dotted] (I31.north) -- ($(I31.north)!1.0!(c1.south)$);
    \draw[-] (I32.north) -- (c3.south);
    \draw[-] (c3.south) -- (I43.north);
    \draw[-] (c3.south) -- ($(c3.south)!0.8!(I53.north)$);
    \draw[dotted] (c3.south) -- ($(c3.south)!1.0!(I53.north)$);
    \draw[-] (I42.north) -- (c4.south);
    \draw[-] (I43.north) -- (c4.south);
    \draw[-] (c4.south) -- ($(c4.south)!0.4!(I54.north)$);
    \draw[dotted] (c4.south) -- ($(c4.south)!0.8!(I54.north)$);
    \draw[-] (c4.south) -- ($(c4.south)!0.28!(I64.north)$);
    \draw[dotted] (c4.south) -- ($(c4.south)!0.6!(I64.north)$);

\end{tikzpicture}
    \caption{Factor graph representation of \eqref{eq:ubfgeq_factorization} for $L=2$.}
    \label{fig:UBFGEQ}
\end{figure}
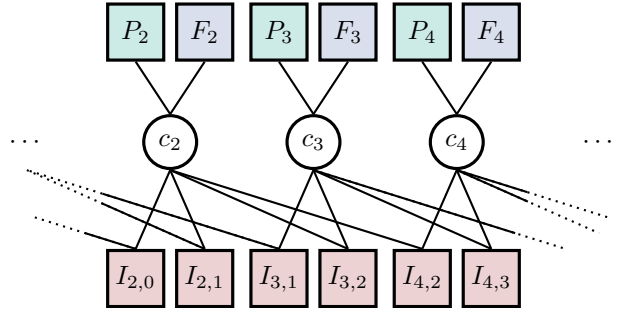

The computation of the symbol-wise \acp{APP} $P(\rv{c}_k = c | \vek{y})$ in \eqref{eq:symbol_wise_app} 
requires $K$ marginalizations which we can efficiently compute by employing the factor graph framework. 
To model a factor graph, we need to find an appropriate factorization of the \ac{APP} distribution $P(\vek{c} | \vek{y})$.
The likelihood in \eqref{eq:likelihood} can be expressed as~\cite{colavolpe_siso_2011}
\begin{align*}
    p(\vek{y|c}) &\propto \exp\mleft( {2{\text{Re}}\mleft\{ \vek{c}^{\textrm H} \vek{H}^{\textrm H} \vek{y} \mright\} -\vek{c}^{\textrm H}\vek{H}^{\textrm H}\vek{H}\vek{c}\over \sigma^2}  \mright).
\end{align*}
We substitute
\begin{equation}
    \vek{G} := \vek{H}^{\textrm H} \vek{H}, \qquad
    \vek{x} := \vek{H}^{\textrm H} \vek{y} \label{eq:x_matched_filter}
\end{equation}
and interpret $\vek{x}$ as an alternative observation at the receiver.
This is commonly known as the \emph{Ungerboeck observation model}~\cite{ungerboeck_adaptive_1974}. 
By using
\begin{align*}
    \vek{c}^{\textrm H}\vek{x} &= \sum\limits_{k=1}^{K} x_k c_k^{\star} \\
    \vek{c}^{\textrm H}\vek{G}\vek{c} &= \sum\limits_{k=1}^{K} G_{k,k} |c_{k}|^2 - \sum\limits_{k=1}^{K} \sum\limits_{\substack{\ell=1 \\ \ell \neq k}}^{K} \text{Re} \mleft\{ G_{k,\ell} c_{\ell} c_{k} \mright\},
\end{align*}
the likelihood function can be factorized as
\begin{equation*}
    p(\vek{y|c}) \propto \prod\limits_{k=1}^{K} \left [ {F_k(c_k) }
        \prod\limits_{\substack{\ell=0 \\ \ell \neq k}}  {J_{k,\ell}(c_k,c_\ell) } 
        \right ]
\end{equation*}
with the factors
\begin{align}
    F_k(c_k) &:=  \exp \mleft( \frac{1}{\sigma^2} {\text{Re}} \mleft\{ 2 x_k c_k^\star - G_{k,k}|c_k|^2 \mright\} \mright) \label{eq:f_fn} \\
    J_{k,\ell}(c_k,c_{\ell}) &= \exp \mleft(  -\frac{1}{\sigma^2} {\text{Re}} \mleft\{  G_{k,\ell} c_\ell c_k^\star  \mright \}  \mright). \label{eq:j_fn}
\end{align}
The factors $J_{k,\ell}(c_k, c_\ell)$ and $J_{\ell,k}(c_{\ell}, c_k)$ depend on the same variables and can thus be condensed to one factor
\begin{align}
    I_{k,\ell}(c_k,c_\ell) :={}& J_{k,\ell}(c_k, c_\ell)  J_{\ell,k}(c_\ell, c_k), \quad k > \ell \label{eq:I_summary} \\
    ={}& \color{cR3}{J^2_{k,\ell}(c_k, c_\ell),} \label{eq:I_simplification}
\end{align}
where \eqref{eq:I_simplification} exploits the Hermitian symmetry of $\vek{G}$.
The factor $I_{k,\ell}(c_k,c_\ell)$ is symmetric with respect to $k$ and $\ell$,
i.e., $I_{k,\ell}(c_k,c_\ell) = I_{\ell,k}(c_\ell,c_k)$.
The a priori distribution 
\begin{equation*}
    P(\vek{c}) = \prod_{k=1}^{K} P(c_k)
\end{equation*}
can be factorized due to the statistical independence of the information symbols.
In summary, the \ac{APP} can be expressed in the factorization
\begin{equation} \label{eq:ubfgeq_factorization}
    P(\vek{c}|\vek{y}) \propto 
    \prod_{k=1}^{K} P(c_k)
    \prod\limits_{k=1}^{K} \left [ {F_k(c_k) }
    \prod\limits_{\ell<k} {I_{k,\ell}(c_k,c_\ell) }
    \right ],
\end{equation}
which is represented by a factor graph in Fig.~\ref{fig:UBFGEQ}.
Applying the \ac{SPA} on this factor graph to develop a symbol detection algorithm was first proposed by Colavolpe et al.\ in~\cite{colavolpe_siso_2011}. We will refer to this algorithm as UFG (Ungerboeck-based factor graph symbol detector) in the following.
We initialize all messages with ${\mu_0(c_k):=-\ln\mleft(M\mright)}$ and perform $N$ \ac{SPA} iterations on the graph. We apply a flooding schedule, i.e., one iteration comprises the simultaneous update of all messages from \acp{VN} to \acp{FN} in a first step, followed by the update of messages propagating in the opposite direction in a second step. The soft output $\hat{P}(c_k|\vek{y})$ is finally obtained by applying~\eqref{eq:loopy_marginal} to all \acp{VN}.

The complexity of factor graph-based algorithms can be estimated by considering the number of \acp{FN} and their node degree, since the \ac{FN} update rule~\eqref{eq:f2v_update} is computationally more demanding than the operation~\eqref{eq:v2f_update} at the \acp{VN}~\cite{colavolpe_application_2005}.
The UFG symbol detector is based on a factor graph with maximum \ac{FN} degree of $2$ of the $I_{k,\ell}$ nodes. 
Therefore, the algorithm has a computational complexity which only grows linearly with the channel memory $L$.
This makes the UFG algorithm an attractive low-complexity alternative to the well established BCJR algorithm which has an exponentially growing complexity with $L$.

Although the factorization \eqref{eq:ubfgeq_factorization} is exact, the UFG algorithm only yields an approximation for the symbol-wise \acp{APP}
due to cycles within the underlying factor graph. It is thus a suboptimal algorithm.
By agglomerating variable nodes, the cycles within the factor graph can be eliminated.
This leads to a forward-backward algorithm described in~\cite{colavolpe_map_2005}, yielding the exact marginalization $P(c_k|\vek{y})$, on which
optimum \ac{MAP} detection can be carried out. 
However, the appealing complexity properties of the cyclic factor
graph vanish if clustering is applied: the \ac{SPA} algorithm on the clustered factor graph has a complexity similar to the BCJR algorithm~\cite{colavolpe_map_2005}, which grows exponentially in both channel memory $L$ and number of bits per symbol $m$.

\section{Neural Enhancement of Factor Graph-Based Symbol Detection}\label{sec:neural_enhancement}
Driven by the appealing complexity properties of the UFG algorithm, we
urge to compensate for its suboptimality by neurally enhancing both the factor graph and the \ac{SPA}.
In particular, we consider \ac{NBP} and an optimization of the \acp{FN} in Sec.~\ref{subsec:nbp}.
We further propose a dynamic factor graph transition in Sec.~\ref{subsec:graph_transition}, specifically tailored to this particular problem.
Based thereupon, we present a novel symbol detection algorithm, which is formally defined in Sec.~\ref{sec:alg}.
Section~\ref{subsec:opt} details the parameter optimization
and introduces the \ac{BMI} as an objective function for optimization and evaluation.

\subsection{Neural Belief Propagation and FN Enhancement}\label{subsec:nbp}
\begin{figure}[tb]
    \centering
    \tikzstyle{fn} = [draw, very thick, regular polygon, regular polygon sides=4, minimum width = 2.25em, inner sep=0pt]
\tikzstyle{fn_ghost} = [very thick, regular polygon, regular polygon sides=4, minimum width = 2.25em, inner sep=0pt]
\tikzstyle{vn} = [draw, very thick, circle, inner sep=0pt, minimum size = 1.5em]
\tikzstyle{vn_ghost} = [very thick, circle, inner sep=0pt, minimum size = 1.5em]

\begin{tikzpicture}[auto, node distance=4em and 4em, thick]
\clip (1.7, -5.7) rectangle (9.2, 1.0);
    \node [vn_ghost] (c0){};
    \node [vn_ghost, right= of c0] (c1){$\cdots$};
    \node [vn, label=center:$c_{2}$, right= of c1] (c2){};
    \node [vn, label=center:$c_{3}$, right= of c2] (c3){};
    \node [vn, label=center:$c_{4}$, right= of c3] (c4){};
    \node [vn_ghost, right = of c4] (c5){};
    \node [right=1em of c4](c_dots){$\cdots$};
    \node [vn_ghost, right= of c5] (c6){};
    
    \node [fill=KITgreen, fill opacity=0.2, fn, label=center: $P_{2}$, left=0.5em of c2.170, anchor = south east] (p2){};
    \node [fill=KITgreen, fill opacity=0.2, fn, label=center: $P_{3}$, left=0.5em of c3.170, anchor = south east] (p3){};
    \node [fill=KITgreen, fill opacity=0.2, fn, label=center: $P_{4}$, left=0.5em of c4.170, anchor = south east] (p4){};
    \node [fill=KITblue, fill opacity=0.2, fn, label=center: $F_{2}$, left=0.5em of c2.190, anchor = north east] (f2){};
    \node [fill=KITblue, fill opacity=0.2, fn, label=center: $F_{3}$, left=0.5em of c3.190, anchor = north east] (f3){};
    \node [fill=KITblue, fill opacity=0.2, fn, label=center: $F_{4}$, left=0.5em of c4.190, anchor = north east] (f4){};

    \node [vn_ghost, below = 11em of c0] (cc0){};
    \node [vn_ghost, right= of cc0] (cc1){$\cdots$};
    \node [vn, label=center:$c_{2}$, right= of cc1] (cc2){};
    \node [vn, label=center:$c_{3}$, right= of cc2] (cc3){};
    \node [vn, label=center:$c_{4}$, right= of cc3] (cc4){};
    \node [vn_ghost, right = of cc4] (cc5){};
    \node [right=1em of cc4](cc_dots){$\cdots$};
    \node [vn_ghost, right= of cc5] (cc6){};
    
    \node [fill=KITgreen, fill opacity=0.2, fn, label=center: $P_{2}$, left=0.5em of cc2.170, anchor = south east] (pp2){};
    \node [fill=KITgreen, fill opacity=0.2, fn, label=center: $P_{3}$, left=0.5em of cc3.170, anchor = south east] (pp3){};
    \node [fill=KITgreen, fill opacity=0.2, fn, label=center: $P_{4}$, left=0.5em of cc4.170, anchor = south east] (pp4){};
    \node [fill=KITblue, fill opacity=0.2, fn, label=center: $F_{2}$, left=0.5em of cc2.190, anchor = north east] (ff2){};
    \node [fill=KITblue, fill opacity=0.2, fn, label=center: $F_{3}$, left=0.5em of cc3.190, anchor = north east] (ff3){};
    \node [fill=KITblue, fill opacity=0.2, fn, label=center: $F_{4}$, left=0.5em of cc4.190, anchor = north east] (ff4){};
    
    \node [fill=KITred, fill opacity=0.2, fn, label=center: $I_{2,0}$, left=0em of $(c2.260)!0.5!(cc2.100)$] (I20){};
    \node [fill=KITred, fill opacity=0.2, fn, label=center: $I_{2,1}$, right=0em of $(c2.280)!0.5!(cc2.80)$] (I21){};
    \node [fill=KITred, fill opacity=0.2, fn, label=center: $I_{3,1}$, left=0em of $(c3.260)!0.5!(cc3.100)$] (I31){};
    \node [fill=KITred, fill opacity=0.2, fn, label=center: $I_{3,2}$, right=0em of $(c3.280)!0.5!(cc3.80)$] (I32){};
    \node [fill=KITred, fill opacity=0.2, fn, label=center: $I_{4,2}$, left=0em of $(c4.260)!0.5!(cc4.100)$] (I42){};
    \node [fill=KITred, fill opacity=0.2, fn, label=center: $I_{4,3}$, right=0em of $(c4.280)!0.5!(cc4.80)$] (I43){};
    \node [fn_ghost, left=0em of $(c5.260)!0.5!(cc5.100)$] (I53){};
    \node [fn_ghost, right=0em of $(c5.280)!0.5!(cc5.80)$] (I54){};
    \node [fn_ghost, left=0em of $(c6.260)!0.5!(cc6.100)$] (I64){};
    \node [fn_ghost, right=0em of $(c6.280)!0.5!(cc6.80)$] (I65){};
    
    \node [vn_ghost, above = 10em of c0] (cp0){};
    \node [vn_ghost, right= of cp0] (cp1){};
    \node [vn_ghost, right= of cp1] (cp2){};
    \node [vn_ghost, right= of cp2] (cp3){};
    \node [vn_ghost, right= of cp3] (cp4){};
    \node [vn_ghost, right = of cp4] (cp5){};
    \node [vn_ghost, right= of cp5] (cp6){};
    
    \node [fn_ghost, left=0em of $(c2.260)!0.5!(cp2.100)$] (Ip20){};
    \node [fn_ghost, right=0em of $(c2.280)!0.5!(cp2.80)$] (Ip21){};
    \node [fn_ghost, left=0em of $(c3.260)!0.5!(cp3.100)$] (Ip31){};
    \node [fn_ghost, right=0em of $(c3.280)!0.5!(cp3.80)$] (Ip32){};
    \node [fn_ghost, left=0em of $(c4.260)!0.5!(cp4.100)$] (Ip42){};
    \node [fn_ghost, right=0em of $(c4.280)!0.5!(cp4.80)$] (Ip43){};
    \node [fn_ghost, left=0em of $(c5.260)!0.5!(cp5.100)$] (Ip53){};
    \node [fn_ghost, right=0em of $(c5.280)!0.5!(cp5.80)$] (Ip54){};
    \node [fn_ghost, left=0em of $(c6.260)!0.5!(cp6.100)$] (Ip64){};
    \node [fn_ghost, right=0em of $(c6.280)!0.5!(cp6.80)$] (Ip65){};
    
    \node [vn_ghost, below = 11em of cc0] (cs0){};
    \node [vn_ghost, right= of cs0] (cs1){};
    \node [vn_ghost, right= of cs1] (cs2){};
    \node [vn_ghost, right= of cs2] (cs3){};
    \node [vn_ghost, right= of cs3] (cs4){};
    \node [vn_ghost, right = of cs4] (cs5){};
    \node [vn_ghost, right= of cs5] (cs6){};
    
    \node [fn_ghost, left=0em of $(cc2.260)!0.5!(cs2.100)$] (Is20){};
    \node [fn_ghost, right=0em of $(cc2.280)!0.5!(cs2.80)$] (Is21){};
    \node [fn_ghost, left=0em of $(cc3.260)!0.5!(cs3.100)$] (Is31){};
    \node [fn_ghost, right=0em of $(cc3.280)!0.5!(cs3.80)$] (Is32){};
    \node [fn_ghost, left=0em of $(cc4.260)!0.5!(cs4.100)$] (Is42){};
    \node [fn_ghost, right=0em of $(cc4.280)!0.5!(cs4.80)$] (Is43){};
    \node [fn_ghost, left=0em of $(cc5.260)!0.5!(cs5.100)$] (Is53){};
    \node [fn_ghost, right=0em of $(cc5.280)!0.5!(cs5.80)$] (Is54){};
    \node [fn_ghost, left=0em of $(cc6.260)!0.5!(cs6.100)$] (Is64){};
    \node [fn_ghost, right=0em of $(cc6.280)!0.5!(cs6.80)$] (Is65){};
    
    \draw[-] (c2.west) -- (p2.east);
    \draw[-] (c2.west) -- (f2.east);
    \draw[-] (c3.west) -- (p3.east);
    \draw[-] (c3.west) -- (f3.east);
    \draw[-] (c4.west) -- (p4.east);
    \draw[-] (c4.west) -- (f4.east);
    \draw[-] (cc2.west) -- (pp2.east);
    \draw[-] (cc2.west) -- (ff2.east);
    \draw[-] (cc3.west) -- (pp3.east);
    \draw[-] (cc3.west) -- (ff3.east);
    \draw[-] (cc4.west) -- (pp4.east);
    \draw[-] (cc4.west) -- (ff4.east);
    
    \draw[dotted] (I20.north) -- ($(I20.north)!0.4!(c0.south)$);
    \draw[latex-] (I20.north) -- ($(I20.north)!0.2!(c0.south)$);
    \draw[dotted] (c1.south) -- (I31.north);
    \draw[latex-] (I31.north) -- ($(I31.north)!0.7!(c1.south)$);
    \draw[-latex, dotted] (c1.south) -- (I21.north);
    \draw[latex-] (I21.north) -- ($(I21.north)!0.58!(c1.south)$);
    \draw[-latex] (c2.south) -- (I42.north);
    \draw[-latex] (c2.south) -- (I32.north);
    \draw[-latex] (c2.south) -- (I20.north);
    \draw[-latex] (c2.south) -- (I21.north);
    \draw[dotted] (c3.south) -- (I53.north);
    \draw[-] (c3.south) -- ($(c3.south)!0.8!(I53.north)$);
    \draw[-latex] (c3.south) -- (I43.north);
    \draw[-latex] (c3.south) -- (I32.north);
    \draw[-latex] (c3.south) -- (I31.north);
    \draw[dotted] (c4.south) -- (I64.north);
    \draw[-] (c4.south) -- ($(c4.south)!0.28!(I64.north)$);
    \draw[dotted] (c4.south) -- (I54.north);
    \draw[-] (c4.south) -- ($(c4.south)!0.42!(I54.north)$);
    \draw[-latex] (c4.south) -- (I43.north);
    \draw[-latex] (c4.south) -- (I42.north);
    
    \draw[-latex, dotted] (cc2.south) -- (Is42.north);
    \draw[-latex, dotted] (cc2.south) -- (Is32.north);
    \draw[-latex, dotted] (cc2.south) -- (Is20.north);
    \draw[-latex, dotted] (cc2.south) -- (Is21.north);
    \draw[-latex, dotted] (cc3.south) -- (Is53.north);
    \draw[-latex, dotted] (cc3.south) -- (Is43.north);
    \draw[-latex, dotted] (cc3.south) -- (Is32.north);
    \draw[-latex, dotted] (cc3.south) -- (Is31.north);
    \draw[-latex, dotted] (cc4.south) -- (Is64.north);
    \draw[-latex, dotted] (cc4.south) -- (Is54.north);
    \draw[-latex, dotted] (cc4.south) -- (Is43.north);
    \draw[-latex, dotted] (cc4.south) -- (Is42.north);
    
    \draw[-latex] (I20.south) -- (cc2.north);
    \draw[dotted] (I20.south) -- ($(I20.south)!0.4!(cc0.north)$);
    \draw[-] (I20.south) -- ($(I20.south)!0.2!(cc0.north)$);
    \draw[-latex] (I21.south) -- (cc2.north);
    \draw[-] (I21.south) -- ($(I21.south)!0.58!(cc1.north)$);
    \draw[dotted] (I21.south) -- (cc1.north);
    \draw[-latex] (I31.south) -- (cc3.north);
    \draw[dotted] (I31.south) -- (cc1.north);
    \draw[-] (I31.south) -- ($(I31.south)!0.7!(cc1.north)$);
    \draw[-latex] (I32.south) -- (cc3.north);
    \draw[-latex] (I32.south) -- (cc2.north);
    \draw[-latex] (I42.south) -- (cc4.north);
    \draw[-latex] (I42.south) -- (cc2.north);
    \draw[-latex] (I43.south) -- (cc4.north);
    \draw[-latex] (I43.south) -- (cc3.north);
    \draw[dotted] (I53.south) -- (cc3.north);
    \draw[latex-] (cc3.north) -- ($(cc3.north)!0.8!(I53.south)$);
    \draw[dotted] (I54.south) -- (cc4.north);
    \draw[latex-] (cc4.north) -- ($(cc4.north)!0.42!(I54.south)$);
    \draw[dotted] (I64.south) -- (cc4.north);
    \draw[latex-] (cc4.north) -- ($(cc4.north)!0.28!(I64.south)$);
    
    \node[left=0em of $(c1)!0.5!(cc1)$, rotate=-90, anchor=center](ln){\small Layer $1$};
    \node[below=-0.5em of cc1.south, rotate=-90, anchor=west](ln+1){\small Layer $2$};
\end{tikzpicture}
    \caption{Unrolled \ac{SPA} on the factor graph of Fig.~\ref{fig:UBFGEQ}: each layer corresponds to one iteration of the \ac{SPA}.}
    \label{fig:NBP_UBFGEQ}
\end{figure}
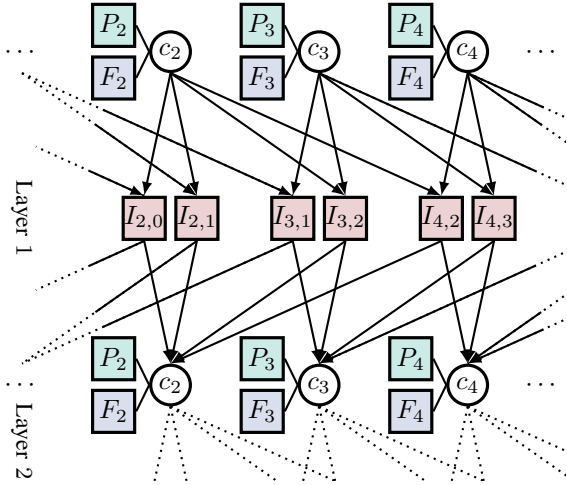

Applying the \ac{SPA} to cyclic factor graphs yields an iterative algorithm.
By the use of deep unfolding, first introduced in~\cite{gregor_learning_2010},
an iterative algorithm can be converted into a \ac{DNN}.
If a flooding schedule is applied to the factor graph in Fig.~\ref{fig:UBFGEQ}, 
a single iteration of the \ac{SPA} consists of propagating messages from \acp{VN}
to \acp{FN} and back.
Unfolding the $N$ iterations of the \ac{SPA} on the factor graph is thus natural since each iteration 
is already (factor) graph-based.
The resulting unrolled network comprises $N$ layers and is shown in Fig.~\ref{fig:NBP_UBFGEQ}. 
Messages are propagated through the \ac{DNN} in a feed-forward fashion.
For the sake of simplicity, we introduce a shorter notation for the messages in the \ac{DNN}:
\begin{align*}
    \mu_{k,j}^{(n)}(c_k) :={}& \mu_{c_k \rightarrow I_{k,k+j}}^{(n)}(c_k) \\
    \nu_{k,j}^{(n)}(c_k) :={}& \mu_{I_{k,k+j} \rightarrow c_k}^{(n)}(c_k).
\end{align*}
Due to the inherent band structure of $\vek{G}$, the network is not fully connected and the index $j$ is limited to ${j \in \mathcal{J} := \{ -L,\ldots,-1,1,\ldots,L\}}$.
\acp{VN} and \acp{FN} accept incoming messages from the previous layer and apply the \ac{SPA} message update rule. The resulting outgoing messages are forwarded downstream to the next layer.
As a consequence, each transmitted message in every iteration has its individually assigned edge.
By accordingly weighting each message/edge, we attempt to mitigate the effects of short cycles 
and improve the detection performance compared to the UFG algorithm. 
Optimizing the weights towards a loss function
by the use of established deep learning techniques is known as \ac{NBP}, 
first introduced in the context of \ac{BP} on Tanner graphs for decoding of linear block codes~\cite{nachmani_learning_2016}.

We limit the set of weights to the edges between \acp{VN} $c_k$ and \acp{FN} $I_{k,\ell}$. 
The \acp{FN} $P_k$ and $F_k$ have degree~$1$. Incident edges are thus not included in the cycles of the factor graph and are consequently not weighted.
In consistency with the message notation, the message $\mu_{k,j}^{(n)}(c_k)$ is multiplied by the weight $w_{\text{v},k,j}^{(n)}$.
Vice versa, messages from \acp{FN} to \acp{VN} $\nu_{k,j}^{(n)}(c_k)$ are weighted by
$w_{\text{f},k,j}^{(n)}$. 
We further generalize the \acp{FN} of the UFG algorithm by the application of multiplicative weights $\kappa_{i,k}^{(n)}$ and $\lambda_{k,\ell}^{(n)}$ within the \ac{FN} computation, in order to increase the parameter optimization space further.
The generalized factors, given in the logarithmic domain, are
\begin{align*}
    \tilde{F}^{(n)}_k(c_k) &:=\frac{\kappa^{(n)}_{1,k}}{\sigma^2} {\text{Re}} \mleft\{ {\kappa^{(n)}_{2,k}} 2 x_k c_k^* - {\kappa^{(n)}_{3,k}} G_{k,k}|c_k|^2 \mright\} \label{eq:F_gen} \\
    \tilde{I}^{(n)}_{k,\ell}(c_k,c_\ell) &:= \lambda^{(n)}_{k,\ell} 
    \mleft( \tilde{J}_{k,\ell}(c_k, c_\ell) +  \tilde{J}_{\ell,k}(c_\ell, c_k) \mright)\\
    \tilde{J}_{k,\ell}(c_k,c_{\ell}) &:= \ln \mleft( J_{k,\ell}(c_k,c_{\ell}) \mright) = -\frac{1}{\sigma^2} {\text{Re}} \mleft\{  G_{k,\ell} c_\ell c_k^\star  \mright\}
    .
\end{align*}
One central motivation to introduce the (trainable) weights $\lambda_{k,\ell}^{(n)}$ and $\kappa_{1,k}^{(n)}$ is to artificially attenuate the $1/\sigma^2$ term inside the \acp{FN}. By adopting a value of $1/\sigma^2$ in the factor graph smaller than the actual one, the overconfidence of the \ac{SPA} messages can be reduced, by describing the channel as if it added more noise than it actually does~\cite{colavolpe_siso_2011}.
In summary, the set of parameters for the generalized algorithm is
\begin{align*}
    \mathcal{P}_{\text{NBP}} := \Big\{ &w_{\text{v},k,j}^{(n)}, w_{\text{f},k,j}^{(n)}, \kappa_{i,k}^{(n)}, \lambda_{k,\ell}^{(n)}, \, n = 1,\ldots, N, \\ 
    &j \in \mathcal{J}, \, i = 1,2,3, \, k = 1,\ldots, K \Big\}
\end{align*}
and contains $|\mathcal{P}_{\text{NBP}}| = NK(5L+3)$ real-valued elements.
Note that we define all parameters of $\mathcal{P}_\text{NBP}$ to be independent of $c_k$. For instance, we constrain the weights $w_{\text{v},k,j}^{(n)}$ and $w_{\text{f},k,j}^{(n)}$ to be scalars although the \ac{SPA} messages $\mu_{k,j}^{(n)}(c_k)$ and $\nu_{k,j}^{(n)}(c_k)$ are $M$-dimensional vectors. This limitation significantly reduces the total number of parameters which need to be optimized.
The UFG algorithm is a special instance of the proposed generalization and is obtained by the parametrization $\mathcal{P}_{\text{NBP}} = \mathcal{P}_1 := \{1,\ldots,1 \}$.
By optimizing $\mathcal{P}_{\text{NBP}}$, the performance of the resulting
algorithm can thus not be inferior to the UFG algorithm, but might yield a performance gain~\cite{nachmani_learning_2016}. 

Note that our approach of directly enhancing the UFG algorithm by generalizing its underlying graph and weighting the \ac{SPA} messages is conceptually different from neural augmentation techniques such as \acp{GNN}~\cite{satorras_neural_2021}. Neural augmentation does not modify the model-based algorithm directly but instead utilizes an external \ac{DNN} to correct the \ac{SPA} messages of the original algorithm in each iteration~\cite{shlezinger_model-based_2020}.

\subsection{Dynamic Factor Graph Transition}\label{subsec:graph_transition}
One key principle of the \ac{SPA} is the extrinsic information rule.
By computing an outgoing message only based on incoming messages of extrinsic edges, 
the \ac{SPA} ensures that only ``new information'' is propagated through the graph.
The extrinsic concept is violated in a cyclic factor graph, where messages propagate
within a loop repetitively and intrinsic information is mistaken for extrinsic information by the nodes 
involved in the cycle.
This violation is inevitable if a sufficient number of iterations is performed
on a cyclic factor graph. 

In Sec.~\ref{subsec:nbp}, we have introduced a generalization of the \acp{FN} as well
as \ac{NBP} in order to mitigate the performance degradation due to the cycles.
However, inherent cycles still exist in the unfolded architecture of the \ac{NBP} and the proposed methods might not be able to fully compensate for their existence. 
Consequently, we propose an additional
strategy to reduce the effect of cycles in a more intrinsic way. 
By dynamically modifying the underlying factor graph on which the message passing is iteratively
performed, messages do not repeatedly arrive at one and the same factor node because either the
graph structure and/or the \acp{FN} therein have changed.
In specific, we propose to periodically change the factor graph's underlying observation model by the application of a linear filter.
The channel output $\rvvek{y}$ is therefore preprocessed by a \ac{FIR} filter and
the result ${\rvvek{x}=\vek{P} \rvvek{y}}$ is then used as a new observation for the inference task. $\vek{P}$ is a band-structured convolutional Toeplitz matrix based on the generic impulse response ${\vek{p}\in \mathbb{C}^{L_\text{p}+1}}$ of the \ac{FIR} preprocessing filter.

The factorization in \eqref{eq:ubfgeq_factorization} represents the \ac{APP} and
is thus optimal in the context of Bayesian inference.
In order to maintain this optimality, the output of the preprocessor $\rvvek{x}$ must
be a \emph{sufficient statistic} for the estimation of $\rvvek{c}$~\cite{krishnamurthy_sufficient_2015}. 
Only in this case, the data processing inequality holds with equality~\cite[Sec.~2.8]{cover_elements_1991} and the original observation $\rvvek{y}$
is irrelevant for the MAP detection, if $\rvvek{x}$ is available~\cite{colavolpe_siso_2011}.
In fact, the original UFG algorithm implicitly uses a preprocessor in \eqref{eq:x_matched_filter} by applying a matched filter ${\vek{P} = \vek{H}^{\text H}}$  to the channel output $\vek{y}$.
Based on this Ungerboeck observation model, inference is carried out using the new observation ${\vek{x} = \vek{H}^{\text H} \vek{y}}$.  
In order to generate a multitude of different factor graphs,
we generalize the observation model
\begin{equation*}
    \tilde{\vek{x}} := \vek{P}\vek{y}, \qquad
    \tilde{\vek{G}} := \vek{P}\vek{H}.
\end{equation*}
Note that varying $\tilde{\vek{G}}$ directly affects the underlying factor graph in \eqref{eq:f_fn} and \eqref{eq:j_fn}.
The matrix $\tilde{\vek{G}}$ is in general not Hermitian symmetric. 
Consequently, the simplification in \eqref{eq:I_simplification} is not valid and
the factor $I_{k,\ell}$ is defined by \eqref{eq:I_summary}.
This, however, does not change the structure of the underlying
factor graph but only of the \acp{FN} therein.
By using different preprocessors, we gain distinct factor graph instances.
We can leverage multiple factor graph instances in two dimensions:
\begin{itemize}
    \item \textbf{Dynamic factor graph transition}: Instantiate $S$ different factor graphs, so-called \emph{stages}. Perform $N'$ \ac{SPA} iterations on one stage $s$, before proceeding to the next stage $s+1$.
    \item \textbf{Parallelism}: In each stage, apply $B$ distinct factor graph instances in parallel. Combine the $B$ results after each stage $s$ to improve the quality of the \ac{APP} estimation.
\end{itemize}
\begin{figure}[t]
    \centering
    \input{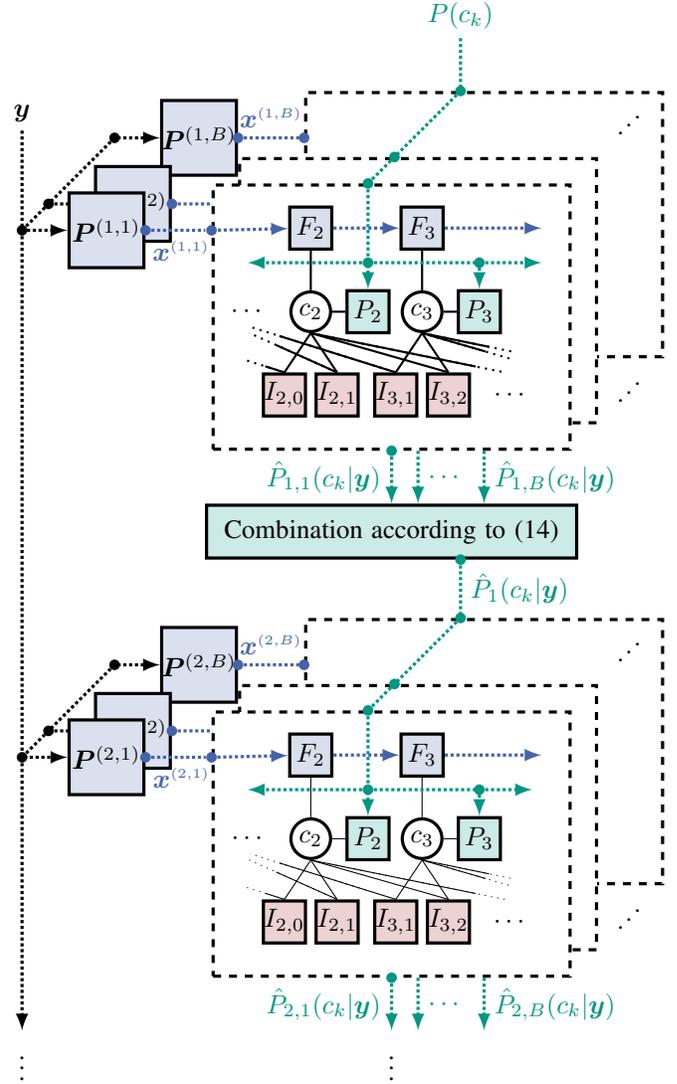}
    \caption{Hierarchical structure of the GAP algorithm with $B$ branches and $S$ stages for a channel with memory ${L=2}$. The GAP algorithm accepts statistical a priori information $P(c_k)$ about the information symbols $c_k$ as well as the channel observation $\vek{y}$. It returns an estimation of the symbol-wise \ac{APP} distributions $\hat{P}_S(c_k|\vek{y})$.}
    \label{fig:GAP_flowgraph}
\end{figure}
Based on this idea, we propose the novel symbol detection algorithm GAP (graph alteration by preprocessing). 
Figure~\ref{fig:GAP_flowgraph} summarizes the information flow of the GAP algorithm. Each of the ${S\cdot B}$ factor graph instances uses its individual preprocessor
${\vek{P}^{(s,b)}, \, s=1,\ldots,S, \, b=1,\ldots,B}$ and thus its individual observation model ${\vek{x}^{(s,b)} = \vek{P}^{(s,b)}\vek{y}}$.
The $B$ factor graphs of a stage $s$ work independently and in parallel. Each factor graph performs $N'$ \ac{SPA} iterations, based on $\vek{x}^{(s,b)}$.
To merge the $B$ individual results $\hat{P}_{s,b}(c_k|\vek{y})$,
we combine the logarithmic \ac{APP} distributions of the symbol detectors by addition, followed by a normalization:
\begin{align}\label{eq:merge_soft_info}
    \ln \mleft( \hat{P}_{s}(c_k|\vek{y}) \mright) =
    &\sum\limits_{b=1}^{B} \ln \mleft( \hat{P}_{s,b}(c_k|\vek{y}) \mright) \nonumber \\
    &- \maxstar\limits_{\text{m} \in \mathcal{M}} \mleft( \sum\limits_{b=1}^{B} \ln \mleft( \hat{P}_{s,b}(c_k = \text{m}|\vek{y}) \mright) \mright).
\end{align}
The idea of using multiple parallel processors of varying behavior in order to increase overall performance was already proposed in the context of channel decoding on Tanner graphs, e.g., in~\cite{hehn_multiple-bases_2010}. Our method, however, differs in the way of how to combine the information of the parallel branches. Instead of selecting the most reliable output of all branches or uniformly averaging over the probabilities, both proposed in~\cite{hehn_multiple-bases_2010}, our approach in \eqref{eq:merge_soft_info} inherently weights the prioritization of the individual results based on the parametrization $\mathcal{P}_\text{NBP}$.

The combined result $\hat{P}_{s}(c_k|\vek{y})$ is passed to the subsequent stage, by setting the \acp{FN} $P_k^{(n)}$ of the factor graphs in stage ${s+1}$ and branch $b$ to
\begin{equation*}
    P_k^{(n)} = \exp \left( w_{\text{p},s,b}^{(n)} \cdot \ln \left(\hat{P}_{s}(c_k|\vek{y})\right) \right), \quad n=1,\ldots,N'.
\end{equation*}
The parameters $w_{\text{p},s,b}^{(n)}$ are introduced to control the dependency between adjacent stages. Note that $\hat{P}_{s}(c_k|\vek{y})$ is only an (imprecise) approximation of the symbol-wise \ac{APP} distributions $P(c_k|\vek{y})$. Finding a more precise approximation is the objective of the subsequent stages, respectively.
Therefore, the weights $w_{\text{p},s,b}^{(n)}$ can dampen the influence of $\hat{P}_{s}(c_k|\vek{y})$, e.g., in the final iterations ${n \in N'}$.
We summarize the introduced parameters for each stage $s$ and branch $b$ in
\begin{equation*}
    \mathcal{P}_\text{p}^{(s,b)}
    := \mleft \{ w_{\text{p},s,b}^{(n)}, \, {n = 1,\ldots,N'} \mright \}.
\end{equation*}
Within a specific factor graph instance with fixed $s$ and $b$, we use the shorter notation ${w_\text{p}^{(n)}}$.

\subsection{Algorithm and Complexity}\label{sec:alg}
\begin{algorithm}[t]
    \DontPrintSemicolon
    \KwData{$\vek{y}$, $P_\text{e}(c_k)$, $\mathcal{P}_\text{p}$,  $\mathcal{P}_\text{NBP}$, $\vek{P}$, $N$}
    Preprocessing $\tilde{\vek{x}} = \vek{P} \vek{y}$ \;
    Initialize messages $\nu_{k,j}^{(0)}(c_k) = -\ln \mleft( M \mright)$ \;
    \For{$n = 1,\ldots,N$} %
    {
        $\mu_{k,j}^{(n)}(c_k) = \!
        w_{\text{v},k,j}^{(n)} \mleft( \xi_k^{(n)}(c_k) + 
        \sum\limits_{{j' \in \mathcal{J} \setminus j } }
        \nu_{k,j'}^{(n-1)}(c_k) \mright)$ \;
        
        $\nu_{k,j}^{(n)}(c_k) = \!
        w_{\text{f},k,j}^{(n)}
        \maxstar\limits_{c_{k+j}} \mleft( 
        \tilde{I}_{k,k+j}^{(n)}(c_k) + 
        \mu_{k+j,-j}^{(n)}(c_k) \mright)$\label{alg:f2v_update}\; 
    }       
    Compute symbol-wise \ac{APP} estimates
    $\hat{P}(c_k|\vek{y}) =
    \exp \mleft( \xi_k^{(N+1)}(c_k) + 
    \sum\limits_{j' \in \mathcal{J}} \nu_{k,j'}^{(N)} (c_k) \mright)$ \;
    \KwResult{$\hat{P}(c_k|\vek{y}), \quad k=1,\ldots,K$}
    \caption{GFG}
    \label{alg:gfg}
\end{algorithm}
We formalize the discussed methods for the neural enhancement of factor graph-based symbol detection.
Algorithm~\ref{alg:gfg} defines the GFG (generalized factor graph-based detection) algorithm which generalizes the UFG algorithm by \ac{NBP}, neurally enhanced \acp{FN} as well as a generalized observation model.
The GFG algorithm is parametrized by $\mathcal{P}_\text{NBP}$, $\mathcal{P}_\text{p}$ and the preprocessor ${\vek{P}\in \mathbb{C}^{(K+2L,K+L)}}$, which is assumed to be a band-structured Toeplitz matrix and describes the convolution with an \ac{FIR} filter ${\vek{p} \in \mathbb{C}^{L_\text{p}+1}}$.
The algorithm accepts the channel output ${\vek{y}\in \mathbb{C}^{K+L}}$ as well as extrinsic information $P_\text{e}(c_k)$ of the information symbols $c_k$, e.g., statistical a priori knowledge ${P_\text{e}(c_k) = P(c_k)}$.
The preprocessor is applied to the channel observation and all messages from
\acp{FN} $I_{k,\ell}(c_k,c_\ell)$ to \acp{VN} $c_k$ are initialized to the same value.
According to the \ac{SPA}, messages of degree~$1$ \acp{FN} do not receive extrinsic information and are consequently not updated. 
The messages from the \acp{FN} $\tilde{F}_k(c_k)$ and $P_k(c_k)$ to the \acp{VN} $c_k$ can thus be computed in advance and are summarized in
\begin{equation*}
    \xi_k^{(n)}(c_k) := w_\text{p}^{(n)} \cdot \ln \mleft( P_\text{e}(c_k) \mright) + \tilde{F}_k^{(n)}(c_k).
\end{equation*}
Subsequently, messages are passed iteratively between \acp{FN} and \acp{VN}
based on the \ac{SPA}.
The message update in line~\ref{alg:f2v_update} is simplified due to the degree~$2$ nature of the \acp{FN} $\tilde{I}_{k,\ell}$.
The symbol-wise \ac{APP} estimates $\hat{P}(c_k|\vek{y})$ are computed by a final marginalization for ${k=1,\ldots,K}$ and are the result of the GFG algorithm.

\begin{algorithm}[t]
    \DontPrintSemicolon
    \KwData{$\vek{y}$, $P(c_k)$, $\mathcal{P}_{\text{GAP}}$}
    Initialize extrinsic information $\hat{P}_0 = P(c_k)$ \;
    \For{$s = 1,\ldots,S$}
    {
        \For{$b = 1,\ldots,B$}
        {
        $\hat{P}_{b,s}(c_k) =
        \text{GFG}_{b,s} \mleft( 
        \vek{y}, \hat{P}_{s-1}(c_k), \mathcal{P}_\text{p}^{(s,b)}, \mathcal{P}_\text{NBP}^{(s,b)}, \vek{P}^{(s,b)}, N'
        \mright)$
        }
        Merge \ac{APP} estimates $\hat{P}_{b,s}(c_k)$ of all branches $b=1,\ldots B$ into $\hat{P}_s(c_k)$ according to \eqref{eq:merge_soft_info}.
    }
    \KwResult{Symbol-wise \ac{APP} estimates $\hat{P}(c_k|\vek{y}) = \hat{P}_S(c_k), \quad k=1,\ldots,K$}
    \caption{GAP ${(S,B,N')}$}
    \label{alg:GAP}
\end{algorithm}
Based on the dynamic factor graph transition discussed in Sec.~\ref{subsec:graph_transition}, we formally define the novel symbol detection algorithm GAP. The GAP algorithm is a hierarchical detection algorithm, structured in $S$ stages and $B$ branches as illustrated in Fig.~\ref{fig:GAP_flowgraph}. Each of the ${B\cdot S}$ units can be seen as an individual GFG symbol detector which is characterized by a unique parametrization $\mathcal{P}_\text{NBP}^{(s,b)}$, $\mathcal{P}_\text{p}^{(s,b)}$ and an individual preprocessor ${\vek{P}^{(s,b)} \in \mathbb{C}^{(K+2L,K+L)}}$. 
The GAP algorithm is defined in Algorithm~\ref{alg:GAP}.
It accepts the channel output $\vek{y}$, statistical a priori knowledge $P(c_k)$ about the information symbols
as well as the parametrization
\begin{equation*}
    \mathcal{P}_\text{GAP} := \bigcup\limits_{s=1}^S \bigcup\limits_{b=1}^B
    \mathcal{P}_{\text{NBP}}^{(s,b)}\cup \mathcal{P}_{\text{p}}^{(s,b)} 
    \cup \mleft \{  \vek{P}^{(s,b)}  \mright \},
\end{equation*}
where ${\mathcal{P}_{\text{NBP}}^{(s,b)}}$ denotes the set ${\mathcal{P}_{\text{NBP}}}$ for each individual GFG unit in stage $s$ and branch $b$.
The parameter set contains $|\mathcal{P}_\text{GAP}|=SB(|\mathcal{P}_{\text{NBP}}|+N+2(L_\text{p}+1))$ real-valued elements.
Note that all GFG units in one stage are fed by the same extrinsic information ${P_\text{e}(c_k)}$ and only differ in their individual parametrization and preprocessor.

Concerning the computational complexity, the GAP algorithm shares the appealing properties of the UFG algorithm.
Due to a constant node degree~$2$ of the \acp{FN} $\tilde{I}_{k,\ell}(c_k)$, the complexity grows linearly with the channel memory $L$, 
and quadratically with the size $M$ of the constellation alphabet~$\mathcal{M}$.
Depending on the number of branches and stages, the GAP algorithm has an order of complexity $O(SBN'KLM^2)$. 
In transmission scenarios over channels with large memory $L$ or high-order constellations, the
GAP algorithm becomes a low-complexity alternative to the BCJR algorithm which has an asymptotic complexity of $O(KM^{L+1})$.
For a detailed complexity comparison, Table~\ref{tab:complexity} reports the number of real additions (ADD), real multiplications (MULT) and $\maxstar$~operations for the algorithms UFG, GAP as well as for the BCJR algorithm. One-time operations for initialization that are independent of the channel observation as well as boundary effects are neglected.
\begin{table*}[t]
    \centering
\caption{Number of operations per information symbol for different symbol detectors operating in the logarithmic domain}
\begin{tabular}{ r l  l  l  l }
  \toprule
    Algorithm & ADD & MULT & max$^\star$\\ %
  \midrule			
    BCJR & $8 M^{L+1}$ & $2M^{L+1}$ &  $3M^{L+1}$ \\
    UFG     &$M(2L+4+N(6L+1))+2L$& $2(L+M+1)$& $2NLM^2$ \\ %
    GAP & $BS [ 2L_{\text{p}}+M[N'(6L_{\text{p}} + 4)+(2L_{\text{p}}+1)+M]+SM$ & $BS ( 2L_{\text{p}}+1) (2N'M + 1)$ & $MS[2BN'LM +1]$ \\ %
  \bottomrule
\end{tabular}
\label{tab:complexity}
\end{table*}
Table~\ref{tab:specific_complexity} evaluates the complexity for a selection of specific transmission scenarios with a given channel memory $L$ and constellation size $M$. The complexity parameter $X$ is defined to be the total number of operations, comprising real additions, multiplications and $\maxstar$ operations that are required to estimate the \ac{APP} of one symbol~$c_k$. Note that the given parametrizations for $(S,B,N')$ are of examplary nature and need to be specifically adapted to different channels in practice.
See Sec.~\ref{sec:results} for realistic parametrizations on some specific channels. 
\begin{table}[tb]
    \centering
    \caption{Complexity parameter $X$ for $N=10$ and different
    configurations of the constellation size $M$ and the channel memory $L$}
\begin{tabular}{l   rr   rr }
    \toprule
    Algorithm & \multicolumn{2}{c}{BPSK $(M=2)$} & \multicolumn{2}{c}{16-QAM $(M=16)$} \\
    \midrule
     &$L=4$ & $L=10$ & $L=4$ & $L=10$  \\
     \textcolor{white}{GAP}~~~$(S,B,N')$         &$L_\text{p}=9$ & $L_\text{p}=10$ & $L_\text{p}=9$ & $L_\text{p}=10$ \\
  \midrule			
    BCJR  & 416 & 26624               & $\approx 10^{7}$ & $\approx 10^{14}$ \\
  UFG & 866 & 2114              & 24722 & 61418\\
  GAP~~~$(1,\,1,\,10)$  & 2723 & 3011            & 62069 & 68825 \\
  GAP~~~$(5,\,2,\,\textcolor{white}{0}4)$  & 11460 & 12660 & 252610 & 279850 \\
  \bottomrule  
\end{tabular}
    \label{tab:specific_complexity}
\end{table}

\subsection{Parameter Optimization}\label{subsec:opt}
The parameters in $\mathcal{P}_\text{GAP}$ are jointly optimized towards an objective function in an end-to-end manner.
Since the GAP algorithm embodies a \ac{NN}, 
we rely on a rich pool of advanced optimization and training methods
developed for feed-forward neural networks in the last years.
For training, we use the Adam algorithm~\cite{kingma_adam_2015}; 
a stochastic gradient descent optimizer.
The gradient can be computed using backpropagation~\cite{rumelhart_learning_1986}, which is a standard method for \acp{NN}.
All weights $\mathcal{P}_\text{NBP}^{(s,b)}$ and $\mathcal{P}_\text{p}^{(s,b)}$ are initialized with $1.0$.
The initial impulse responses $\vek{p}$ of the preprocessors $\vek{P}^{(s,b)}$ are independently sampled from a standard normal distribution.

We optimize the parametrization towards a maximum achievable rate between the channel input and the detector output.
Many practical transmission systems use \ac{BICM}, which decouples the 
symbol detection from a binary soft-decision \ac{FEC}~\cite{Fabregas_foundations_2008}.
In \ac{BICM}, the symbol detector soft output $\hat{P}(c_k|\vek{y})$ is converted by a \ac{BMD} to binary soft information  
\begin{equation*}
    \hat{P}(\rv{b}_i(\rv{c}_k) = b | \vek{y}) = 
    \sum\limits_{c \in \mathcal{M}_i^{(b)}} \hat{P}(\rv{c}_k = c | \vek{y}), \quad b \in \{ 0,1 \}
\end{equation*}
with
$\mathcal{M}_i^{(b)} := \mleft \{ c \in \mathcal{M} : b_{i}(c) = b \mright \}$.
The resulting bit-wise \acp{APP} are typically expressed in \acp{LLR}
\begin{equation*}
    L_{k,i}(\vek{y},\alpha) := \alpha \ln \mleft( \frac{\hat{P}(\rv{b}_i(\rv{c}_k) = 0 | \vek{y})}{\hat{P}(\rv{b}_i(\rv{c}_k) = 1 | \vek{y})} \mright), \quad \alpha\in \mathbb{R}_{> 0}.
\end{equation*}
If the \ac{LLR} is based on suboptimal detection, 
i.e., if ${\hat{P}(\rv{b}_i = b|\vek{y})}$ is not the true \ac{APP}, the scaling factor $\alpha$ corrects a potential \ac{LLR} mismatch~\cite[Chap.~7]{szczecinski_bit-interleaved_2015}.
After interleaving, the \acp{LLR} are fed to a bit-wise soft-decision \ac{FEC}.
By interpreting the \ac{BMD} as a mismatched detector, the \ac{BMI} is an achievable information rate for \ac{BICM}~\cite{Fabregas_foundations_2008}.
The calculation of the \ac{BMI}, detailed in~\cite{alvarado_achievable_2018}, 
considers the \ac{BMD} by assuming $m$ parallel sub-channels 
transmitting on a binary basis
instead of one symbol-based channel. Assuming \ac{iid} transmit bits,
the \ac{BMI}\footnote{Note that the \ac{BMI} is often called \ac{GMI} in literature. \ac{GMI}, however, is a more general concept and defines a lower bound of the mismatched capacity. For the special case of a mismatched decoder due to bit-metric decoding, the \ac{GMI} is equivalent to the \ac{BMI}~\cite{martinez_bit-interleaved_2009}.} 
is defined as the sum of mutual informations
$I(\rv{b}_i ; \rv{y})$ of $m$ unconditional bit-wise channel transmissions:
\begin{align*}
    \text{BMI} :={}& \sum _{i=1}^{m}\! I(\rv{b}_i(\rv{c}_k) ; \rv{y}) \\
    ={}& \sum _{i=1}^{m}\!\mathbb{E}_{\rv{b}_i,\rv{y}}\mleft \{ \log_2 \mleft( \frac{P_{\rv{b}_{i}(\rv{c}_k)|\rv{y}}(\rv{b}_i|\rv{y})}{P_{\rv{b}_i}(\rv{b}_i(\rv{c}_k))} \mright) \mright \}.
\end{align*}
By a sample mean estimation over $D$ labeled data batches
$\mathcal{D} := \{ (\vek{c}^{(\text{d})}, \vek{y}^{(\text{d})})_{i}:
\vek{c}^{(\text{d})} \in \mathcal{M}^K,
\vek{y}^{(\text{d})} = \vek{H} \vek{c}^{(\text{d})} + \vek{w}_i , i=1,\ldots,D \}$
and by assuming uniformly distributed information bits,
a feasible approximation
\begin{align}\label{bmi_est}
    &\text{BMI} \approx \log_2 \mleft( M \mright) - \\
    &\frac{1}{D K} \sum\limits_{\mathclap{i=1}}^{m} 
    \sum\limits_{k=0}^{K-1}
    \sum\limits_{\mathclap{\qquad \quad (\vek{c}^{(\text{d})}, \vek{y}^{(\text{d})})\in \mathcal{D}}} 
    \log_2 \mleft (  \exp\mleft(-(-1)^{b_{k,i}(c_k^{(\text{d})})} L_{k,i}(\vek{y}, \alpha)\mright) + 1  \mright ) \nonumber
\end{align}
can be found~\cite{alvarado_achievable_2018}.
The \ac{BMI} can be used to evaluate the soft decision performance of \ac{SISO} symbol detectors in numerical simulations.
For suboptimal detectors, the optimum $\alpha$ which maximizes the BMI can be determined in an efficient way, e.g., by the Golden section search~\cite{schmalen_performance_2017}.
For gradient descent optimization, we employ the metric in~\eqref{bmi_est} with ${\alpha=1}$ as an objective function.

For the optimization of the GAP algorithm with multiple stages $S>1$, we can increase the gradient update of the backpropagation by using multiloss terms~\cite{nachmani_deep_2018}. Hence, we suggest the term
\begin{align}\label{eq:multiloss}
    \text{BMI}_\text{multi} := &\log_2 \mleft( M \mright) - 
    \frac{1}{S D K} \sum\limits_{\mathclap{s=1}}^{S} \sum\limits_{\mathclap{i=1}}^{m} 
    \sum\limits_{k=0}^{K-1}
    \sum\limits_{\mathclap{\qquad \quad (\vek{c}^{(\text{d})}, \vek{y}^{(\text{d})})\in \mathcal{D}}} \\
    &\log_2 \mleft (  \exp\mleft(-(-1)^{b_{k,i}(c_k^{(\text{d})})} L_{k,i}^{(s)}(\vek{y}, \alpha)\mright) + 1  \mright ) \nonumber
\end{align}
as the average BMI between the transmitted bits and the \acp{LLR} obtained from the \ac{APP} estimates after each stage $s$, which we denote with $L_{k,i}^{(s)}(\vek{y}, \alpha)$.
Using the multiloss term in \eqref{eq:multiloss} for the optimization of the GAP algorithm improves learning of the earlier stages~\cite{nachmani_deep_2018}. Note that we use the default loss \eqref{bmi_est} unless explicitly stated differently.

The parameters $S,B,N'$ and $L_\text{p}$ define the general dimensionality of the model and are not part of the optimization process described in this section. These so called hyperparameters define the overall behavior and complexity of the GAP algorithm and can be either chosen by hand, or optimized in a so-called hyperparameter tuning, as elaborated in~\cite{raschka_model_2020}.

\section{Numerical Results}\label{sec:results}
\begin{table}[tb]
    \centering
\caption{Characterization of linear \ac{ISI} reference channels}
\begin{tabular}{ l l  r  l }
  \toprule
    Name & Alias & $L$ & Impulse Response $\vek{h}$ \\
  \midrule			
    Proakis~A & $\text{C}_\text{A}$ & 10 & $(0.04, -0.05, 0.07, -0.21, -0.5,$ \\
    & & & $ 0.72,0.36, 0.0, 0.21, 0.03, 0.07)^\text{T}$ \\
    Proakis~B & $\text{C}_\text{B}$ & 2 & $(0.407, 0.815, 0.407)^\text{T}$ \\
    Proakis~C & $\text{C}_\text{C}$ & 4 & $(0.227, 0.46, 0.688, 0.46, 0.227)^\text{T}$ \\
  \bottomrule  
\end{tabular}
\label{tab:channels}
\end{table}
We evaluate the considered symbol detectors towards their detection performance.
In all simulations, the information symbols are sampled independently and uniformly from the constellation alphabet $\mathcal{M}$. If not mentioned otherwise, all iterative algorithms perform ${N=10}$ iterations and the trainable parameters are optimized for ${\ebno = 10}$~dB. The source code for the parameter optimization and evaluation of the GAP algorithm is available online~\cite{github_2022}.
The symbol-wise \ac{MAP} detector, implemented by the BCJR algorithm~\cite{bahl_optimal_1974}, as well as a \ac{MMSE} equalizer~\cite[Sec.~9.4]{proakis_digital_2007} with filter order $30$ serve as references. 
For the latter, we transform the soft \ac{MMSE} filter output $\rv{c}_{k,\text{LMMSE}}$ to \acp{APP} ${\hat{P}(c_k|\vek{y})}$ by applying a Gaussian approximation to the estimation error 
$\rv{e}_k := |\rv{c}_{k,\text{LMMSE}} - \rv{c}_k| \sim \mathcal{N}\left(0, \hat{\sigma}^2_\text{LMMSE} \right)$
and estimating the error variance $\hat{\sigma}^2_\text{LMMSE}$ based on the hard decisions $\argmax\limits_{\text{m}\in \mathcal{M}} |\rv{c}_{k,\text{LMMSE}} - \text{m}|$.
We consider a block length of ${K=500}$ symbols and the transmission over three standard \ac{ISI} channel models~\cite[Sec.~9.4]{proakis_digital_2007}, which are characterized in Table~\ref{tab:channels}.

\begin{figure}[tb]
    \centering
    \begin{tikzpicture}[trim axis right]
    \begin{axis}[
          width=\columnwidth, %
          height=0.8\columnwidth,
          grid=major, %
          grid style={gray!30}, %
          xlabel= $E_\text{b}/N_0$ (dB),
          ylabel= BER,
          ymode = log,
          enlarge x limits=false,
          enlarge y limits=false,
          xmin = 0,
          ymax = 0.5,
          ymin = 0.00001,
          legend style={at={(.05,0.05)},anchor=south west, font=\footnotesize},
          legend cell align={left},
          tick label style={font=\footnotesize},
        ]
        \addplot[mark=pentagon*, color=KITblue, line width=1.3pt] table[x=Eb/N0, y=UB_FGEQ_BER ,col sep=comma] {numerical_results/proakisA/BPSK.csv};
        \addplot[dotted, mark=diamond,mark options={scale=2.5,solid}, color=KITblue, line width=1.3pt] table[x=Eb/N0, y=BCJR_BER ,col sep=comma] {numerical_results/proakisA/BPSK.csv};
        \addplot[mark=pentagon*, color=KITpalegreen, line width=1.3pt] table[x=Eb/N0, y=UB_FGEQ_BER ,col sep=comma] {numerical_results/proakisB/BPSK.csv};
        \addplot[dotted, mark=diamond,mark options={scale=2.5,solid}, color=KITpalegreen, line width=1.3pt] table[x=Eb/N0, y=BCJR_BER ,col sep=comma] {numerical_results/proakisB/BPSK.csv};
        \addplot[mark=pentagon*, color=KITred, line width=1.3pt] table[x=Eb/N0, y=UB_FGEQ_BER ,col sep=comma] {numerical_results/proakisC/BPSK.csv};
        \addplot[dotted, mark=diamond,mark options={scale=2.5,solid}, color=KITred, line width=1.3pt] table[x=Eb/N0, y=BCJR_BER ,col sep=comma] {numerical_results/proakisC/BPSK.csv};
        \legend{ $\text{C}_\text{A}$, UFG \\  $\text{C}_\text{A}$, MAP \\ $\text{C}_\text{B}$, UFG \\  $\text{C}_\text{B}$, MAP \\  $\text{C}_\text{C}$, UFG \\  $\text{C}_\text{C}$, MAP \\ }
    \end{axis}
\end{tikzpicture}
    \caption{\ac{BER} over $\ebno$ of the UFG algorithm 
    for a \ac{BPSK} transmission over different linear \ac{ISI} channels.}
    \label{fig:ubfgeq_channels}
\end{figure}
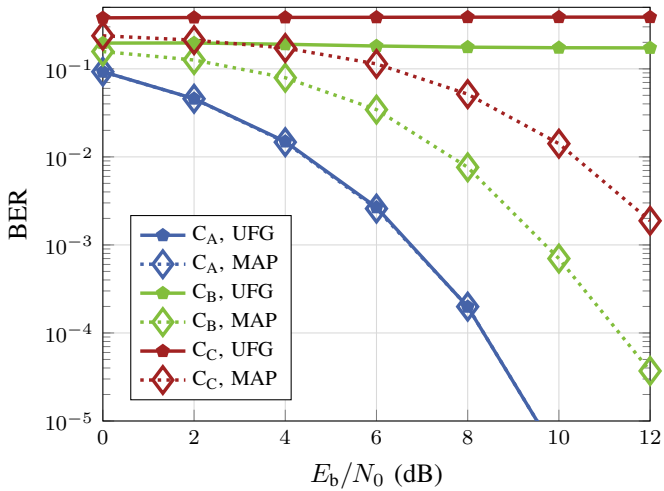
We analyze the detection performance of the original UFG algorithm on all three reference channels.
Note that with the parametrization ${S=B=1}$, ${\vek{P}^{(1,1)} = \vek{H}^{\textrm H}}$, ${\mathcal{P}_{\text{p}}^{(1,1)} = \mathcal{P}_1}$ and ${\mathcal{P}_{\text{NBP}}^{(1,1)} = \mathcal{P}_1}$, the GAP algorithm instantiates the UFG algorithm.
Figure~\ref{fig:ubfgeq_channels} shows the hard-decision performance of the UFG algorithm in terms of the \acf{BER} over $\ebno$ for \acf{BPSK}.
The performance gap to \ac{MAP} detection
is highly channel specific. While the UFG algorithm operates close to optimality for the channel $\text{C}_\text{A}$, its detection capabilities for the channels $\text{C}_\text{B}$ and
$\text{C}_\text{C}$ are quite poor. Notably, the \ac{BER} does not decrease for an increasing $\ebno$.

\subsection{Neural Belief Propagation and FN Enhancement}
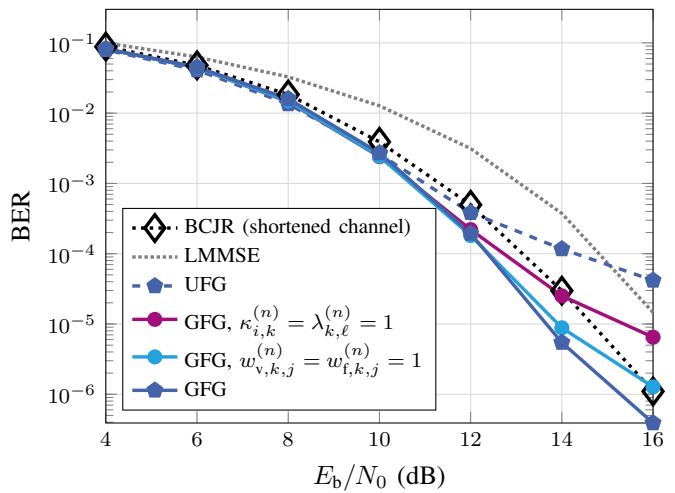
\begin{figure}[tb]
    \centering
    \begin{tikzpicture}[trim axis right]
    \begin{axis}[
          width=\columnwidth, %
          height=0.8\columnwidth,
          grid=major, %
          grid style={gray!30}, %
          xlabel= $E_\text{b}/N_0$ (dB),
          ylabel= BER,
          ymode = log,
          enlarge x limits=false,
          enlarge y limits=false,
          xmin = 4,
          xmax = 16,
          ymax = 0.3,
          legend style={at={(.03,0.03)},anchor=south west, font=\footnotesize},
          legend cell align={left},
          tick label style={font=\footnotesize},
        ]
        \addplot[dotted, mark=diamond,mark options={scale=2.5,solid}, color=black, line width=1.3pt] table[x=Eb/N0, y=short_BCJR_L3N25_BER ,col sep=comma] {numerical_results/proakisA/16QAM.csv};
        \addplot[densely dotted, color=gray, line width=1.3pt] table[x=Eb/N0, y=MMSE_BER ,col sep=comma] {numerical_results/proakisA/16QAM.csv};
        \addplot[dashed, mark=pentagon*,mark options={scale=1.3,solid}, color=KITblue, line width=1.3pt] table[x=Eb/N0, y=UB_FGEQ_BER ,col sep=comma] {numerical_results/proakisA/16QAM.csv};
        
        \addplot[mark=*, KITpurple, line width=1.3pt , mark options={scale=1.1,solid}] table[x=Eb/N0, y=NBP_UB_FGEQ_BER ,col sep=comma] {numerical_results/proakisA/16QAM.csv};
        
        \addplot[mark=*, KITcyan, line width=1.3pt , mark options={scale=1.1,solid}] table[x=Eb/N0, y=UB_FGEQ_SNR*_BER ,col sep=comma] {numerical_results/proakisA/16QAM.csv};
        \addplot[mark=pentagon*, KITblue, line width=1.3pt , mark options={scale=1.3,solid}] table[x=Eb/N0, y=NBP_UB_FGEQ*_BER ,col sep=comma] {numerical_results/proakisA/16QAM.csv};
        
        \legend{BCJR (shortened channel) \\ LMMSE \\ UFG \\ GFG, $\kappa_{i,k}^{(n)} = \lambda_{k,\ell}^{(n)} = 1$ \\ GFG, $w_{\text{v},k,j}^{(n)} = w_{\text{f},k,j}^{(n)} =1$ \\ GFG \\}
        
    \end{axis}
\end{tikzpicture}

\caption{Hard-decision performance of the GFG algorithm with ${\vek{P} = \vek{H}^{\textrm H}}$ and different constraints on $\mathcal{P}_\text{NBP}$ for a 16-QAM transmission over the channel~$\text{C}_\text{A}$. The BCJR algorithm is applied on a shortened channel with an impulse response of length~4 for complexity reasons.}
\label{fig:proakisA}
\vspace{-8pt}
\end{figure}
\begin{figure*}[tb]
\centering
\begin{tabular}{rl}
    
    \begin{tikzpicture}[baseline, trim axis right]
    \begin{axis}[
          width=\columnwidth, %
          height=0.8\columnwidth,
          grid=major, %
          grid style={gray!30}, %
          xlabel= $E_\text{b}/N_0$ (dB),
          ylabel= BMI (bit/channel use),
          enlarge x limits=false,
          enlarge y limits=false,
          xmin = 0,
          ymin = 0.3,
          ymax = 1.0,
          ytick = {0.3, 0.4,0.5,0.6,0.7,0.8,0.9,1.0},
          tick label style={font=\footnotesize},
          legend columns = 2,
          transpose legend,
          legend style={
            /tikz/column 2/.style={column sep=12pt,},
            /tikz/column 4/.style={column sep=12pt,},
            at={(0.34,1.05)},anchor=south west, font=\footnotesize},
            legend cell align={left},
        ]
        \addplot[draw=none, dotted, mark=diamond,mark options={scale=2.5,solid}, color=black, line width=1.3pt] table[x=Eb/N0, y=BCJR_BMI ,col sep=comma] {numerical_results/proakisB/BPSK.csv};
        \addplot[densely dotted, color=gray, line width=1.3pt] table[x=Eb/N0, y=MMSE_BMI ,col sep=comma] {numerical_results/proakisB/BPSK.csv};
        \addplot[dashed, mark=pentagon*,mark options={scale=1.3,solid}, color=KITblue, line width=1.3pt] table[x=Eb/N0, y=UB_FGEQ_BMI ,col sep=comma] {numerical_results/proakisB/BPSK.csv};
        \addplot[mark=pentagon*, KITblue, line width=1.3pt , mark options={scale=1.3,solid}] table[x=Eb/N0, y=NBP_UB_FGEQ*_BMI ,col sep=comma] {numerical_results/proakisB/BPSK.csv};
        \addplot[dashed, mark=square*, mark options={scale=1.3,solid}, color=KITpalegreen, line width=1.3pt] table[x=Eb/N0, y=GFG_P7*_BMI ,col sep=comma] {numerical_results/proakisB/BPSK.csv};
        \addplot[mark=square*, color=KITpalegreen, line width=1.3pt, mark options={scale=1.3,solid}] table[x=Eb/N0, y=NBP_GFG_P7*_BMI ,col sep=comma] {numerical_results/proakisB/BPSK.csv};
        \addplot[dotted, mark=diamond,mark options={scale=2.5,solid}, color=black, line width=1.3pt] table[x=Eb/N0, y=BCJR_BMI ,col sep=comma] {numerical_results/proakisB/BPSK.csv};
        
        \legend{MAP \\ LMMSE \\ UFG \\ GFG, $\vek{P} = \vek{H}^{\textrm H}$ \\ GFG, $\vek{P} =\vek{P}_7^\star$, $\mathcal{P}_\text{NBP} = \mathcal{P}_1$ \\ GFG, $\vek{P} =\vek{P}_7^\star$ \\}
    \end{axis}
\end{tikzpicture}
&
    \begin{tikzpicture}[baseline, trim axis right]
    \begin{axis}[
          width=\columnwidth, %
          height=0.8\columnwidth,
          grid=major, %
          grid style={gray!30}, %
          xlabel= $E_\text{b}/N_0$ (dB),
          ylabel= BER,
          ymode = log,
          enlarge x limits=false,
          enlarge y limits=false,
          xmin = 0,
          ymax = 0.5,
          ymin = 0.00001,
          tick label style={font=\footnotesize},
          legend columns = 2,
          transpose legend,
          legend style={
            /tikz/column 2/.style={column sep=5pt,},
            at={(0.0,1.05)},anchor=south west, font=\footnotesize},
            legend cell align={left},
        ]
        \addplot[dotted, mark=diamond,mark options={scale=2.5,solid}, color=black, line width=1.3pt, draw=none] table[x=Eb/N0, y=BCJR_BER ,col sep=comma] {numerical_results/proakisB/BPSK.csv};
        \addplot[densely dotted, color=gray, line width=1.3pt] table[x=Eb/N0, y=MMSE_BER ,col sep=comma] {numerical_results/proakisB/BPSK.csv};
        \addplot[dashed, mark=pentagon*,mark options={scale=1.3,solid}, color=KITblue, line width=1.3pt] table[x=Eb/N0, y=UB_FGEQ_BER ,col sep=comma] {numerical_results/proakisB/BPSK.csv};
        \addplot[mark=pentagon*, KITblue, line width=1.3pt , mark options={scale=1.3,solid}] table[x=Eb/N0, y=NBP_UB_FGEQ*_BER ,col sep=comma] {numerical_results/proakisB/BPSK.csv};
        \addplot[dashed, mark=square*, mark options={scale=1.3,solid}, color=KITpalegreen, line width=1.3pt] table[x=Eb/N0, y=GFG_P7*_BER ,col sep=comma] {numerical_results/proakisB/BPSK.csv};
        \addplot[mark=square*, color=KITpalegreen, line width=1.3pt, mark options={scale=1.3,solid}] table[x=Eb/N0, y=NBP_GFG_P7*_BER ,col sep=comma] {numerical_results/proakisB/BPSK.csv};
        
        \addplot[dotted, mark=diamond,mark options={scale=2.5,solid}, color=black, line width=1.3pt] table[x=Eb/N0, y=BCJR_BER ,col sep=comma] {numerical_results/proakisB/BPSK.csv};
        
    \end{axis}
\end{tikzpicture}
\end{tabular}
    \caption{\ac{BMI} and \ac{BER} over $\ebno$ for different instances of the GFG algorithm on the  channel $\text{C}_\text{B}$ with \ac{BPSK} signaling.}
    \label{fig:proakisB_BPSK_BER}
\end{figure*}
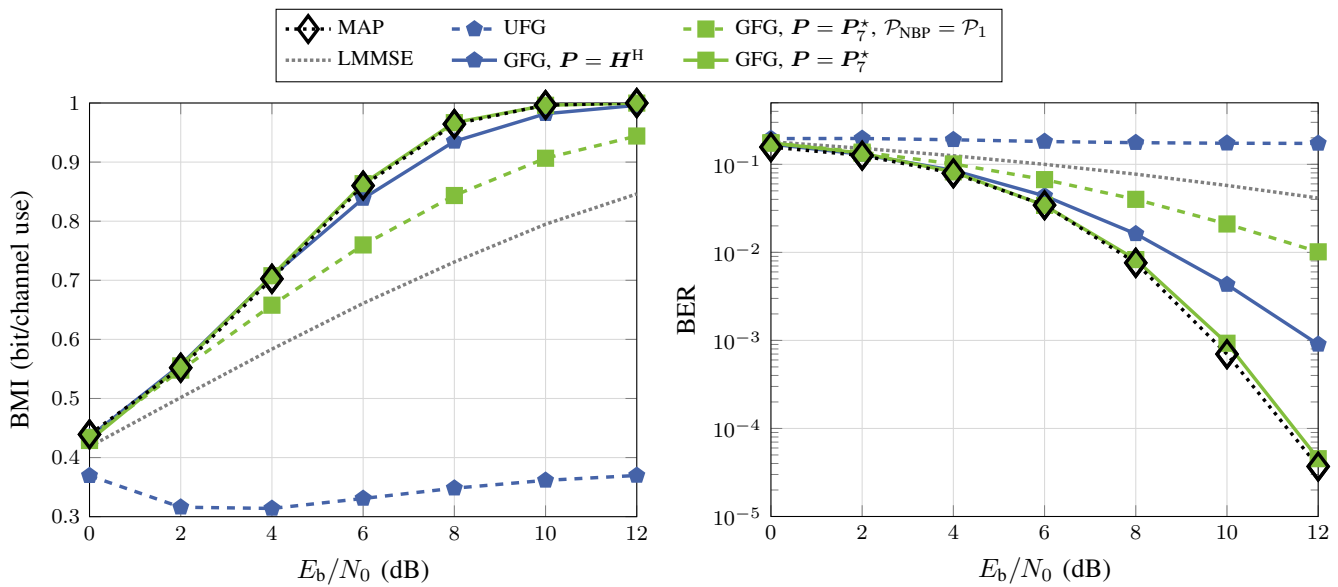
We evaluate the effects of \ac{NBP} and the neural enhancement of the \acp{FN} for the factor graph-based symbol detection on channel $\text{C}_\text{A}$. 
Since the original UFG algorithm already approaches optimum detection performance for \ac{BPSK}, we consider a 16-\ac{QAM} transmission with Gray labeling\footnote{We expect Gray labeling to be optimal w.r.t. the \ac{BER} and \ac{BMI} performance of the proposed detection algorithms.}. Figure~\ref{fig:proakisA} reports the \ac{BER} performance over
$\ebno$.
The UFG algorithm outperforms the \ac{MMSE} equalizer in the low $\ebno$ regime but runs into an error floor.
Applying \ac{NBP} and neurally enhancing the \acp{FN} of the UFG symbol detector, i.e., 
optimizing the parameters $\mathcal{P}_\text{NBP}$ at ${\ebno = 14~\text{dB}}$ for the GFG algorithm with ${\vek{P} = \vek{H}^{\textrm H}}$ mitigates this behavior and generalizes well over the complete $\ebno$ range.
To distinguish between the performance gain due to the weighting of the $\ac{SPA}$ messages, and the generalization of the \acp{FN}, we partly constrain the space of the gradient descent optimization over $\mathcal{P}_\text{NBP}$. First, we fix $\kappa_{i,k}^{(n)} = \lambda_{k,\ell}^{(n)} = 1$, thereby disabling the \ac{FN} generalization. The remaining free parameters are jointly optimized towards the \ac{BMI} which yields a performance improvement. Second, we disable the message weighting by setting $w_{\text{v},k,j}^{(n)} = w_{\text{f},k,j}^{(n)} =1$ and only optimize the remaining parameters within the \acp{FN}. 
Although the dimensionality of the optimization space is about 4 times smaller for the latter case compared to \ac{NBP}, the performance gain is significantly larger. 
However, the best performance is obtained by the combination of both methods and yields a significant performance gain compared to the UFG algorithm as well as the \ac{MMSE} equalizer.
Due to the constellation order ${M=16}$ and a relatively large memory ${L=10}$ of the channel $\text{C}_\text{A}$, \ac{MAP} detection becomes computationally infeasible. In order to nevertheless make a comparison of the proposed algorithm, we filter the received signal $\vek{y}$ with a channel shortening filter and then apply the BCJR algorithm on the shortened channel model. Following~\cite{rusek_optimal_2012}, we can derive an \ac{FIR} channel shortening filter of order~25 which reduces the impulse response length of the channel from 11 (${L=10}$) to 4 (${L=3}$). The detection performance of the BCJR algorithm on the shortened model is reported in Fig.~\ref{fig:proakisA}. It clearly outperforms the \ac{MMSE} equalizer which can be seen as \ac{MAP} detection on a channel model shortened to length~1~\cite{rusek_optimal_2012}. For high $\ebno > 12$~dB, the shortened BCJR algorithm also performs better than the conventional UFG detector, however, it cannot reach the low \acp{BER} of the neurally enhanced GFG algorithm in the considered $\ebno$ range. Note that the complexity of the BCJR algorithm on the shortened channel is still more than ten times higher compared to the proposed GFG detector (complexity parameters ${X=\num{851968}}$ and ${X=\num{68825}}$).

\subsection{Preprocessing}
We examine the sensibility of the factor graph-based GFG algorithm to the observation model. Therefore, we consider symbol detection on the channel $\text{C}_\text{B}$ with \ac{BPSK} signaling in more depth. 
To allow a fair comparison of different observation models,
we initially disable \ac{NBP} and the \ac{FN} generalization for the GFG algorithm by setting ${\mathcal{P}_{\text{NBP}} = \mathcal{P}_1}$. We compare the Ungerboeck observation model with ${\vek{P}=\vek{H}^{\textrm H}}$, which is employed by the UFG algorithm, to a generic preprocessing filter $\vek{P}_7$ of length ${L_\text{p} = 7}$.
The results are given in Fig.~\ref{fig:proakisB_BPSK_BER}.
Optimizing the preprocessor $\vek{P}^\star_7$ with respect to the \ac{BMI}, the symbol detector approaches a \ac{BMI} of $0.9$~bit/channel~use at $\ebno=10$~dB which is a gain of over $0.5$~bit/channel~use compared to the detection based on the Ungerboeck model.

Enabling \ac{NBP} and the \ac{FN} enhancement significantly improves the performance for both considered observation models.
For the detector based on the Ungerboeck model,
the \ac{BER} is thereby reduced by a factor of more than $100$ for ${\ebno = 12}$~dB.
A near-optimum symbol detector is obtained on the channel $\text{C}_\text{B}$ by combining the generalized preprocessing with \ac{NBP} and the \ac{FN} generalization and jointly optimizing all parameters ${\mathcal{P}_\text{NBP}\cup \{ \vek{P}_7\}}$.
Note that the GFG algorithm is a specialization of the GAP algorithm with ${S=B=1}$, i.e., we do not perform a dynamic factor graph transition but only change the observation model once.
An analysis of $\vek{P}_7$ as well of the optimized parameter set $\mathcal{P}_\text{NBP}$ turned out to be not very insightful. Most of the weights approximately follow a Gaussian distribution with mean and variance altering over the iterations $n=1,\ldots,10$. Especially for $\kappa^{(n)}_{1,k}$, it is interesting to observe that its mean is notably smaller than $1.0$ (varying from $0.5$ to $0.9$) for most of the iterations. This supports our hypothesis, discussed in Sec.~\ref{subsec:nbp}, according to which the weights $\kappa^{(n)}_{1,k}$ can attenuate the $1/\sigma^2$ term inside the \acp{FN} of the factor graph, thereby dampening the overconfidence of the \ac{SPA} messages which would otherwise lead to high approximation errors and impairments in the convergence behavior. Only in the last iteration $n=10$, the weights $\kappa^{(n)}_{1,k}$ are amplified with an average  $\kappa^{(10)}_{1,k}$ of $3.1$.

\begin{figure}[tb]
    \centering
    \begin{tikzpicture}[trim axis right]
    
    \begin{axis}[
        name=Ax1,
          ybar = 15pt,
          xbar = 10pt,
          x dir = reverse,
          y dir = reverse,
          bar width=0.25,
          width=0.58\columnwidth,
          height=0.8\columnwidth,
          grid style={gray!30}, %
          xlabel= BER,
          xmode = log,
          ylabel= iteration $n$,
          enlarge x limits=false,
          enlarge y limits=false,
          xmin = 0.0008,
          xmax = 1.0,
          ymin = 0.3,
          ymax = 10.7,
          ytick = {1,2,3,4,5,6,7,8,9,10},
          ytick pos=left,
          xtick = {1.0, 0.1, 0.01, 0.001},
          xticklabels={$10^0$, $10^{-1}$,$10^{-2}$, },
          xtick pos=lower,
          grid=both,
          legend style={at={(0.0,1.03)},anchor=south west, font=\footnotesize},
          legend cell align={left},
          tick label style={font=\footnotesize},
        ]
        \addplot[fill, fill opacity = 0.5, line join=round, dashed,
        mark options={scale=0.7,solid}, color=KITblue, line width=1.3pt, bar shift = 3.5pt] table[x =UBFGEQ, y=iter ,col sep=comma] {numerical_results/proakisB/convergence_proB.csv};
        
        \addplot[fill, fill opacity = 1.0, line join=round,
        mark options={scale=0.7,solid}, color=KITblue, line width=1.3pt, bar shift = -3.5pt] table[x =UFG, y=iter ,col sep=comma] {numerical_results/proakisB/convergence_proB.csv};
        
        \legend{UFG \\ GFG, $\vek{P} = \vek{H}^{\textrm H}$ \\}
        
    \end{axis}
    
        \begin{axis}[
        name=Ax2,
        at={($(Ax1.north east)+(5pt,0)$)},
        anchor=north west,
          ybar=1.5pt,
          xbar=0pt,
          x dir = reverse,
          y dir = reverse,
          bar width=0.25,
          width=0.58\columnwidth,
          height=0.8\columnwidth,
          grid style={gray!30}, %
          xlabel= BER,
          xmode = log,
          ylabel= ,
          enlarge x limits=false,
          enlarge y limits=false,
          xmin = 0.0008,
          xmax = 1.0,
          ymin = 0.3,
          ymax = 10.7,
          ytick = {1,2,3,4,5,6,7,8,9,10},
          ytick style={draw=none},
          yticklabels = {,,,,,,,,,},
          ytick pos=left,
          xtick = {1.0, 0.1,0.01, 0.001},
          xticklabels={$10^0$, $10^{-1}$, $10^{-2}$, $10^{-3}$},
          xtick pos=lower,
          grid=both,
          legend style={at={(1.0,1.03)},anchor=south east, font=\footnotesize},
          legend cell align={left},
          tick label style={font=\footnotesize},
        ]
        \addplot[fill, fill opacity = 0.5, line join=round, dashed,
        mark options={scale=0.7,solid}, color=KITpalegreen, line width=1.3pt, bar shift=3.5pt] table[x =GFG_onlyP, y=iter ,col sep=comma] {numerical_results/proakisB/convergence_proB.csv};
        \addplot[fill, fill opacity = 1.0, line join=round,
        mark options={scale=0.7,solid}, color=KITpalegreen, line width=1.3pt, bar shift =-3.5pt] table[x =GFG, y=iter ,col sep=comma] {numerical_results/proakisB/convergence_proB.csv};
        
        \legend{GFG, $\vek{P} =\vek{P}_7^\star$, $\mathcal{P}_\text{NBP} = \mathcal{P}_1$ \\ GFG, $\vek{P} =\vek{P}_7^\star$ \\}
        
    \end{axis} 

\end{tikzpicture}
    \caption{\ac{BER} over the iterations $n$ of the GFG algorithm on the channel $\text{C}_\text{B}$ at $\ebno = 10$~dB and BPSK signaling.}
    \label{fig:convergence_proB}
\end{figure}
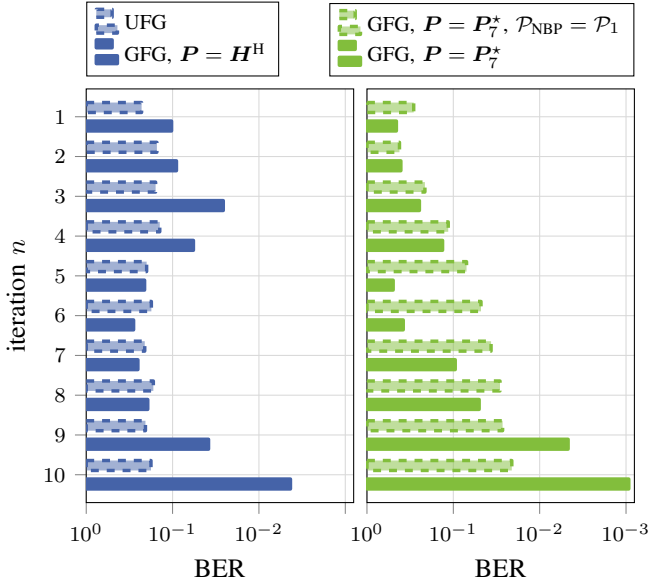
To evaluate the convergence behavior of the considered algorithms, Fig.~\ref{fig:convergence_proB} illustrates the evolution of the \ac{BER} over the \ac{SPA} iterations $n$ for the channel $\text{C}_\text{B}$ at ${\ebno = 10}$~dB. We can observe a non-convergent behavior for the UFG algorithm. The \ac{BER} keeps oscillating between two points at $\text{BER}=0.17$ and $\text{BER}=0.2$ for $n>4$. We can identify this lack of convergence as a major reason for the poor overall performance of the UFG algorithm. Modifying the observation model with the preprocessor $\vek{P} =\vek{P}_7^\star$ fixes this issue and leads to a monotone \ac{BER} convergence for the GFG detector with $\mathcal{P}_\text{NBP} = \mathcal{P}_1$. Applying \ac{NBP} as well as the \ac{FN} optimization to the GFG algorithm leads to a very interesting behavior: for both observation models, $\vek{P} = \vek{H}^{\textrm H}$ and $\vek{P} =\vek{P}_7^\star$, the \ac{BER} first decreases over the iterations, then it climbs again to a local maximum before it reaches its global minimum in the final iteration $n=10$.

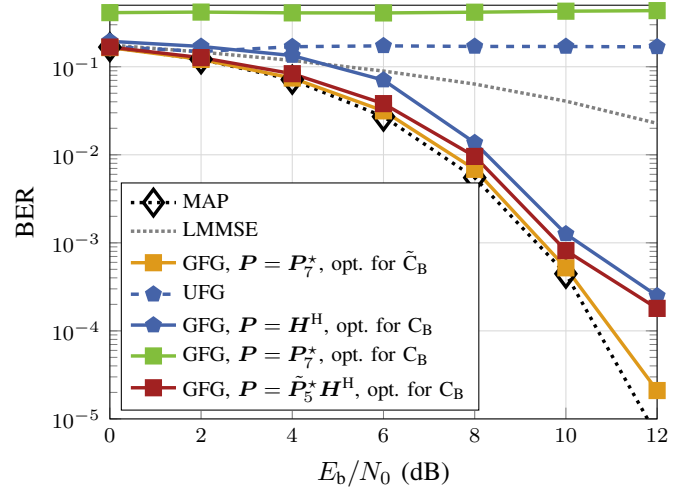
\begin{figure}[tb]
\centering
    \begin{tikzpicture}[trim axis right]
    \begin{axis}[
          width=\columnwidth, %
          height=0.8\columnwidth,
          grid=major, %
          grid style={gray!30}, %
          xlabel= $E_\text{b}/N_0$ (dB),
          ylabel= BER,
          ymode = log,
          enlarge x limits=false,
          enlarge y limits=false,
          ymin = 0.00001,
          ymax = 0.5,
          xmax = 12,
          legend style={at={(0.02, 0.02)},anchor=south west, font=\footnotesize},
          legend cell align={left},
          tick label style={font=\footnotesize},
        ]
        \addplot[dotted, mark=diamond,mark options={scale=2.5,solid}, color=black, line width=1.3pt] table[x=Eb/N0, y=BCJR_BER ,col sep=comma] {numerical_results/altB/altB.csv};
        \addplot[densely dotted, color=gray, line width=1.3pt] table[x=Eb/N0, y=MMSE_BER ,col sep=comma] {numerical_results/altB/altB.csv};
        \addplot[mark=square*, color=KITorange, line width=1.3pt, mark options={scale=1.3,solid}] table[x=Eb/N0, y=GFG_altB_BER ,col sep=comma] {numerical_results/altB/altB.csv};
        \addplot[dashed, mark=pentagon*,mark options={scale=1.3,solid}, color=KITblue, line width=1.3pt] table[x=Eb/N0, y=UBFGEQ_BER ,col sep=comma] {numerical_results/altB/altB.csv};
        \addplot[mark=pentagon*,mark options={scale=1.3,solid}, color=KITblue, line width=1.3pt] table[x=Eb/N0, y=UFG_proB_BER ,col sep=comma] {numerical_results/altB/altB.csv};
        \addplot[mark=square*, color=KITpalegreen, line width=1.3pt, mark options={scale=1.3,solid}] table[x=Eb/N0, y=GFG_proB_BER ,col sep=comma] {numerical_results/altB/altB.csv};
        \addplot[mark=square*, color=KITred, line width=1.3pt, mark options={scale=1.3,solid}] table[x=Eb/N0, y=GFG_var_training_BER ,col sep=comma] {numerical_results/altB/altB.csv};

        \legend{MAP \\ LMMSE \\ 
        GFG, $\vek{P} =\vek{P}_7^\star$, opt.\ for $\tilde{\text{C}}_\text{B}$ \\
        UFG \\ 
        GFG, $\vek{P} = \vek{H}^{\textrm H}$, opt.\ for $\text{C}_\text{B}$ \\
        GFG, $\vek{P} =\vek{P}_7^\star$, opt.\ for $\text{C}_\text{B}$ \\
        GFG, $\vek{P} =\tilde{\vek{P}}_5^\star \vek{H}^{\textrm H}$, opt. for $\text{C}_\text{B}$ \\}
    \end{axis}
    \end{tikzpicture}
    \caption{\ac{BER} performance of the GFG algorithm for a BPSK transmission on the channel $\tilde{\text{C}}_\text{B}$. The parameters of the various GFG instances were optimized for different channels.}
    \label{fig:altB}
\end{figure}

The impressive results of Fig.~\ref{fig:proakisB_BPSK_BER} raise the question of how universal the learned solutions are %
towards variations of the channel. Therefore, we evaluate the instances of the GFG algorithm which were specifically trained on the channel $\text{C}_\text{B}$ at $\ebno=10$~dB (see Fig.~\ref{fig:proakisB_BPSK_BER}) on the alternative channel $\tilde{\text{C}}_\text{B} := (0.59, 0.76, 0.28)^\text{T}$. The channel $\tilde{\text{C}}_\text{B}$ was generated by adding independent and $\mathcal{N}(0,0.1)$-distributed samples to the taps of the impulse response ${\vek{h}_\text{C}}_\text{B}$ of the channel $\text{C}_\text{B}$ and normalizing the result with respect to the channel energy. Figure~\ref{fig:altB} compares the results to an instance of the GFG algorithm which was specifically optimized for the alternative channel $\tilde{\text{C}}_\text{B}$.
Note that all GFG instances still have perfect channel knowledge and only their parametrization is optimized for a ``mismatched'' channel.
We can observe that the neurally enhanced detector GFG with $\vek{P} = \vek{H}^{\textrm H}$ generalizes very well. Although being trained on the channel $\text{C}_\text{B}$, the performance degradation is relatively small compared to the GFG instance which was directly trained on the channel $\tilde{\text{C}}_\text{B}$. In contrast, the GFG algorithm with the optimized preprocessor $\vek{P}=\vek{P}_7^\star$ does not generalize at all. This makes sense, because the factor nodes are entirely based on the mismatched preprocessor $\vek{P}_7^\star$ and do not consider the true channel at all. To fix this, we impose the special structure ${\vek{P} = \tilde{\vek{P}} \vek{H}^{\textrm H}}$ on the preprocessor and only optimize $\tilde{\vek{P}}$ for the channel $\text{C}_\text{B}$ while $\vek{H}^{\textrm H}$ is a matched filter based on the actual channel state information. Evaluating the optimized detector on the alternative channel $\tilde{\text{C}}_\text{B}$ shows that the GFG algorithm also performs well on channels which (slightly) differ from the channel for which the algorithm was optimized, under the condition that the detector has access to perfect channel state information and if the discussed structure of the preprocessor is used.

\subsection{Dynamic Factor Graph Transition}
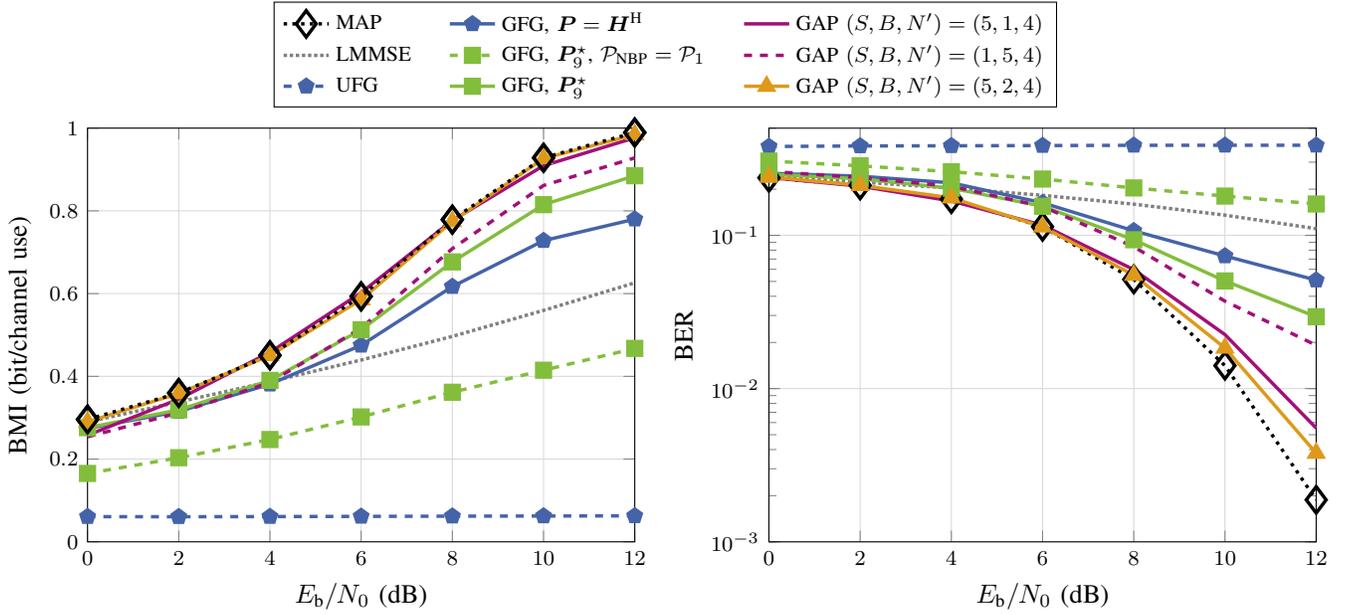
\begin{figure*}[tb]
\centering
\begin{tabular}{rl}
    \begin{tikzpicture}[baseline, trim axis right]
    \begin{axis}[
          width=\columnwidth,
          height=0.8\columnwidth,
          grid=major, %
          grid style={gray!30}, %
          xlabel= $E_\text{b}/N_0$ (dB),
          ylabel= BMI (bit/channel use),
          enlarge x limits=false,
          enlarge y limits=false,
          xmin = 0,
          ymin = 0.0,
          ymax = 1.0,
          tick label style={font=\footnotesize},
          legend columns = 3,
          transpose legend,
          legend style={
            /tikz/column 2/.style={column sep=12pt,},
            /tikz/column 4/.style={column sep=12pt,},
            at={(0.34,1.05)},anchor=south west, font=\footnotesize},
            legend cell align={left},
        ]
        \addplot[draw=none, dotted, mark=diamond,mark options={scale=2.5,solid}, color=black, line width=1.3pt] table[x=Eb/N0, y=BCJR_BMI ,col sep=comma] {numerical_results/proakisC/BPSK.csv};
        \addlegendentry{MAP}
        \addplot[densely dotted, color=gray, line width=1.3pt] table[x=Eb/N0, y=MMSE_BMI ,col sep=comma] {numerical_results/proakisC/BPSK.csv};
        \addlegendentry{LMMSE}
        \addplot[dashed, mark=pentagon*,mark options={scale=1.3,solid}, color=KITblue, line width=1.3pt] table[x=Eb/N0, y=UB_FGEQ_BMI ,col sep=comma] {numerical_results/proakisC/BPSK.csv};
        \addlegendentry{UFG}
        \addplot[mark=pentagon*, color=KITblue, line width=1.3pt, mark options={scale=1.3,solid}] table[x=Eb/N0, y=NBP_UB_FGEQ*_BMI ,col sep=comma] {numerical_results/proakisC/BPSK.csv};
        \addlegendentry{GFG, $\vek{P} = \vek{H}^{\textrm H}$}
        \addplot[dashed, mark=square*,mark options={scale=1.3,solid}, color=KITpalegreen, line width=1.3pt] table[x=Eb/N0, y=GFG_P7*_BMI ,col sep=comma] {numerical_results/proakisC/BPSK.csv};
        \addlegendentry{GFG, $\vek{P}_9^\star$, $\mathcal{P}_\text{NBP} = \mathcal{P}_1$}
        \addplot[mark=square*, color=KITpalegreen, line width=1.3pt, mark options={scale=1.3,solid}] table[x=Eb/N0, y=NBP_GFG_P9*_BMI ,col sep=comma] {numerical_results/proakisC/BPSK.csv};
        \addlegendentry{GFG, $\vek{P}_9^\star$}
        \addplot[mark options={scale=1.5,solid}, color=KITpurple, line width=1.3pt] table[x=Eb/N0, y=GFG2D_514_BMI ,col sep=comma] {numerical_results/proakisC/BPSK.csv};
        \addlegendentry{GAP ${(S,B,N')=(5,1,4)}$}
        \addplot[dashed, mark options={scale=1.5,solid}, color=KITpurple, line width=1.3pt] table[x=Eb/N0, y=GFG2D_154_BMI_s ,col sep=comma] {numerical_results/proakisC/BPSK.csv};
        \addlegendentry{GAP ${(S,B,N')=(1,5,4)}$}
        \addplot[mark=triangle*,mark options={scale=1.5,solid}, color=KITorange, line width=1.3pt] table[x=Eb/N0, y=GFG2D_524_BMI ,col sep=comma] {numerical_results/proakisC/BPSK.csv};
        \addlegendentry{GAP ${(S,B,N')=(5,2,4)}$}
        \addplot[dotted, mark=diamond,mark options={scale=2.5,solid}, color=black, line width=1.3pt] table[x=Eb/N0, y=BCJR_BMI ,col sep=comma] {numerical_results/proakisC/BPSK.csv};
        
    \end{axis}
    \end{tikzpicture}
&
    \begin{tikzpicture}[baseline, trim axis right]
    \begin{axis}[
          width=\columnwidth,
          height=0.8\columnwidth,
          grid=major, %
          grid style={gray!30}, %
          xlabel= $E_\text{b}/N_0$ (dB),
          ylabel= BER,
          ymode= log,
          enlarge x limits=false,
          enlarge y limits=false,
          xmin = 0,
          ymax = 0.5,
          ymin = 0.001,
          legend style={at={(0.05, 0.05)},anchor=south west, font=\footnotesize},
          legend cell align={left},
          tick label style={font=\footnotesize},
        ]
        \addplot[dotted, mark=diamond,mark options={scale=2.5,solid}, color=black, line width=1.3pt] table[x=Eb/N0, y=BCJR_BER ,col sep=comma] {numerical_results/proakisC/BPSK.csv};
        \addplot[densely dotted, color=gray, line width=1.3pt] table[x=Eb/N0, y=MMSE_BER ,col sep=comma] {numerical_results/proakisC/BPSK.csv};
        \addplot[dashed, mark=pentagon*,mark options={scale=1.3,solid}, color=KITblue, line width=1.3pt] table[x=Eb/N0, y=UB_FGEQ_BER ,col sep=comma] {numerical_results/proakisC/BPSK.csv};
        \addplot[mark=pentagon*, color=KITblue, line width=1.3pt, mark options={scale=1.3,solid}] table[x=Eb/N0, y=NBP_UB_FGEQ*_BER ,col sep=comma] {numerical_results/proakisC/BPSK.csv};
        \addplot[dashed, mark=square*,mark options={scale=1.3,solid}, color=KITpalegreen, line width=1.3pt] table[x=Eb/N0, y=GFG_P7*_BER ,col sep=comma] {numerical_results/proakisC/BPSK.csv};
        \addplot[mark=square*, color=KITpalegreen, line width=1.3pt, mark options={scale=1.3,solid}] table[x=Eb/N0, y=NBP_GFG_P9*_BER ,col sep=comma] {numerical_results/proakisC/BPSK.csv};
        \addplot[mark options={scale=1.5,solid}, color=KITpurple, line width=1.3pt] table[x=Eb/N0, y=GFG2D_514_BER ,col sep=comma] {numerical_results/proakisC/BPSK.csv};
        \addplot[dashed, mark options={scale=1.5,solid}, color=KITpurple, line width=1.3pt] table[x=Eb/N0, y=GFG2D_154_BER ,col sep=comma] {numerical_results/proakisC/BPSK.csv};
        \addplot[mark=triangle*,mark options={scale=1.5,solid}, color=KITorange, line width=1.3pt] table[x=Eb/N0, y=GFG2D_524_BER ,col sep=comma] {numerical_results/proakisC/BPSK.csv};
        
    \end{axis}
    \end{tikzpicture}
\end{tabular}
    \caption{Performance of different symbol detectors for a \ac{BPSK} transmission over the channel $\text{C}_\text{C}$. 
    All ${B \cdot S = 8}$ embedded GFG units of the GAP algorithm have an individual preprocessor of length ${L_\text{p}=9}$.}
    \label{fig:proakisC_BPSK_BER}
\end{figure*}
Figure~\ref{fig:proakisC_BPSK_BER} evaluates the performance of different symbol detection algorithms for the channel $\text{C}_\text{C}$ and \ac{BPSK} signaling. 
Applying an extended preprocessor of length ${L_\text{p}=9}$ with optimized filter taps ${\vek{P} = \vek{P}_9^\star}$ to the GFG algorithm can improve the detection performance compared to the UFG algorithm.
However, without the application of \ac{NBP} and the neural \ac{FN} enhancement (${\mathcal{P}_\text{NBP} = \mathcal{P}_1}$), the factor graph-based symbol detection does not outperform the \ac{MMSE} equalizer
but has an approximately constant \ac{BMI} offset of about $0.13$~bit/channel~use.
The application of \ac{NBP} and the neural \ac{FN} enhancement can further improve the GFG algorithm for both the Ungerboeck model ${\vek{P} = \vek{H}^\text{H}}$ and the enhanced preprocessing ${\vek{P} = \vek{P}_9^\star}$, but a significant gap to \ac{MAP} performance remains.
We close this gap to optimal symbol detection by the GAP algorithm.
Applying a dynamic factor graph transition with ${S=5}$ stages and ${B=2}$ parallel branches drastically improves the detection performance. 
Based on the motivation for the dynamic factor graph transition in Sec.~\ref{subsec:graph_transition}, it might seem optimal to constantly alter the factor nodes after
each iteration, i.e., to set ${N' =1}$. In this case, however, we experienced that the convergence of the messages is impaired due to the (too) fast variation of the factor nodes. For the channels which are considered in this work, we found an empirical sweet spot for ${N'=4}$.
To evaluate on the effectiveness of the dynamic factor graph transition and the effect of parallel branches, we compare the performance of the GAP algorithm with ${(S,B,N')=(5,1,4)}$ (i.e., no parallelism ${B=1}$ and ${S=5}$ stages) to the performance of an alternative parametrization with only one stage ${S=1}$ but ${B=5}$ parallel branches. Note that both instances of the GAP algorithm employ the same number of GFG elements which leads to a comparable complexity. The performance evaluation in Fig.~\ref{fig:proakisC_BPSK_BER} reveals the effectiveness of the dynamic factor graph transition: the GAP algorithm with multiple serial stages shows a notably superior performance compared to the GAP algorithm with (only) parallel branches over the entire $\ebno$ range $0-12$~dB considered.
However, the conjunction of both effects, the dynamic factor graph transition (${S=5}$) and the parallelism (${B=2}$), leads to the best overall performance. For the latter parametrization, the overall computational complexity is approximately quadrupled compared to the parametrization ${(S,B,N')=(1,1,10)}$ which instantiates a single GFG element.

\begin{figure}[tb]
    \centering
    \begin{tikzpicture}[trim axis left, trim axis right]
    
    \begin{axis}[
          ybar=1.5pt,
          xbar=0pt,
          y dir = reverse,
          bar width=0.3,
          width=0.95\columnwidth, %
          height=0.9\columnwidth,
          grid style={gray!30}, %
          xlabel= BMI (bit/channel use),
          ylabel= stage $s$,
          enlarge x limits=false,
          enlarge y limits=false,
          xmin = -1,
          xmax = 1,
          ymin = 0.3,
          ymax = 26,
          bar shift=0pt,
          ytick = {5,10,15,20,25},
          yticklabels = {1,2,3,4,5},
          ytick pos=right,
          xtick = {-1,-0.75,-0.5,-0.25,0.0,0.25,0.5,0.75,1.0},
          xticklabels={1,0.75,0.5,0.25,0,0.25,0.5,0.75,1},
          xtick pos=lower,
          grid=both,
          legend style={at={(.05,0.95)},anchor=north west, font=\footnotesize},
          legend cell align={left},
          tick label style={font=\footnotesize},
        ]
        
        \addplot[fill, mark options={scale=1.0,solid}, color=KITorange, line width=0.0pt, bar width=0.75] table[x =stages_BMI, y=s_axis ,col sep=comma] {numerical_results/proakisC/GFG2D_loss_over_stages.csv};
        \addplot[fill, mark options={scale=1.0,solid}, color=KITorange, line width=0.0pt, bar width=0.75] table[x expr={-\thisrow{stages_BMI}}, y=s_axis ,col sep=comma] {numerical_results/proakisC/GFG2D_loss_over_stages.csv};
    \end{axis}

    \begin{axis}[
          ybar=1.5pt,
          xbar=0pt,
          y dir = reverse,
          bar width=0.4,
          width=0.95\columnwidth, %
          height=0.9\columnwidth,
          grid style={gray!30}, %
          xlabel= BMI (bit/channel use),
          ylabel= iteration $n$,
          enlarge x limits=false,
          enlarge y limits=false,
          xmin = -1,
          xmax = 1,
          ymin = 0.3,
          ymax = 26,
          bar shift=0pt,
          ytick = {0,1,2,3,4,5,6,7,8,9,10,11,12,13,14,15,16,17,18,19,20,21,22,23,24},
          yticklabels = {,1,2,3,4,,1,2,3,4,,1,2,3,4,,1,2,3,4,,1,2,3,4},
          ytick pos=left,
          xtick = {-1,-0.75,-0.5,-0.25,0.0,0.25,0.5,0.75,1.0},
          xticklabels={1,0.75,0.5,0.25,0,0.25,0.5,0.75,1},
          xtick pos=lower,
          legend style={at={(0.5,1.03)},anchor=south,legend columns=-1, font=\footnotesize},
          legend cell align={left},
          tick label style={font=\footnotesize},
        ]
        \addplot[fill, fill opacity = 0.5, line join=round,
        mark options={scale=0.7,solid}, color=KITblue, line width=1.3pt] table[x=1_BMI, y=x_axis,col sep=comma] {numerical_results/proakisC/GFG2D_loss_over_iters.csv};
        
        \addplot[fill, fill opacity = 0.5, line join=round, mark options={scale=0.7,solid}, color=KITred, line width=1.3pt] table[x expr={-\thisrow{2_BMI}}, y=x_axis ,col sep=comma] {numerical_results/proakisC/GFG2D_loss_over_iters.csv};
        
        \addplot[fill, mark options={scale=1.0,solid}, color=KITorange, line width=0.0pt, bar width=0.75] table[x =stages_BMI, y=s_axis ,col sep=comma] {numerical_results/proakisC/GFG2D_loss_over_stages.csv};
        
        \legend{$\text{BMI}(\hat{P}^{(n)}_{s,1}(c_k|\vek{y}))$,
        $\text{BMI}(\hat{P}^{(n)}_{s,2}(c_k|\vek{y}))$,
        $\text{BMI}(\hat{P}_s(c_k|\vek{y}))$};
    \end{axis}

\end{tikzpicture}
    \caption{Convergence behavior of the GAP algorithm with ${(S,B,N')=(5,2,4)}$ and ${L_\text{p}=9}$ for a \ac{BPSK} transmission over the channel $\text{C}_\text{C}$. The convergence is analyzed w.r.t.\ the BMI over the $S=5$ stages (orange bars) as well as regarding the $N'=4$ iterations within each stage, separately for both $B=2$ branches (red and blue bars).}
    \label{fig:proC_internal}
\end{figure}
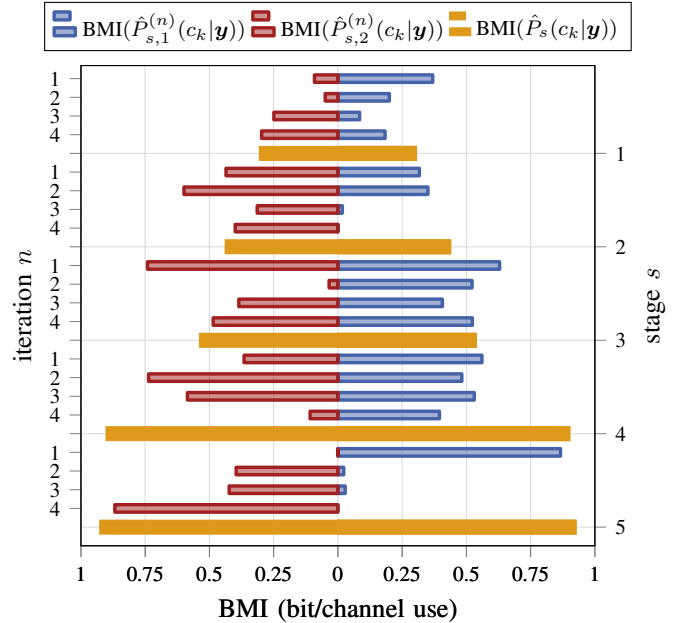
We investigate the behavior of the GAP algorithm with ${(S,B,N')=(5,2,4)}$ and ${L_\text{p} = 9}$ in more depth to gain insights into the effectiveness of the dynamic factor graph transition.
Therefore, we analyze the performance  
after each stage $s$ by approximating the \ac{BMI} based on the interim \ac{APP} estimates
$\hat{P}_s(c_k|\vek{y})$.
The results are reported in Fig.~\ref{fig:proC_internal} (orange bars)
and shows a monotonic increase of the \ac{BMI} over $s$.
To compare the contribution of both branches, we
additionally determine the \ac{BMI} for each branch and iteration individually,
based on $\hat{P}_{s,b}(c_k|\vek{y})$.
Intermediate \ac{BMI} estimates for ${n<N'}$ are obtained by an early termination of the iterative message passing in Algorithm~\ref{alg:gfg} and are denoted by $\hat{P}_{s,b}^{(n)}(c_k|\vek{y})$. Figure~\ref{fig:proC_internal} shows the results for branch ${b=1}$ on the left (red bars) and for branch ${b=2}$ on the right (blue bars). The \ac{BMI} evolution of the single GFG units over the iterations $n$ is highly non-monotonic. Moreover, the two branches have a very distinct behavior.
Especially after the stages ${s=2}$ and ${s=5}$, the GFG output ${\hat{P}_{s,1}(c_k|\vek{y}) = \hat{P}_{s,1}^{(N')}(c_k|\vek{y})}$ of branch ${b=1}$ has a BMI close to zero. However, the combination of both branches, i.e., the combination of ${\hat{P}_{s,1}(c_k|\vek{y})}$ and $\hat{P}_{s,2}(c_k|\vek{y})$ to ${\hat{P}_{s}(c_k|\vek{y})}$ (orange bars), still yields an improved BMI compared to the BMI of one branch $\hat{P}_{s,2}(c_k|\vek{y})$.
This non-intuitive behavior is caused by the optimization process, which is carried out in an end-to-end manner.

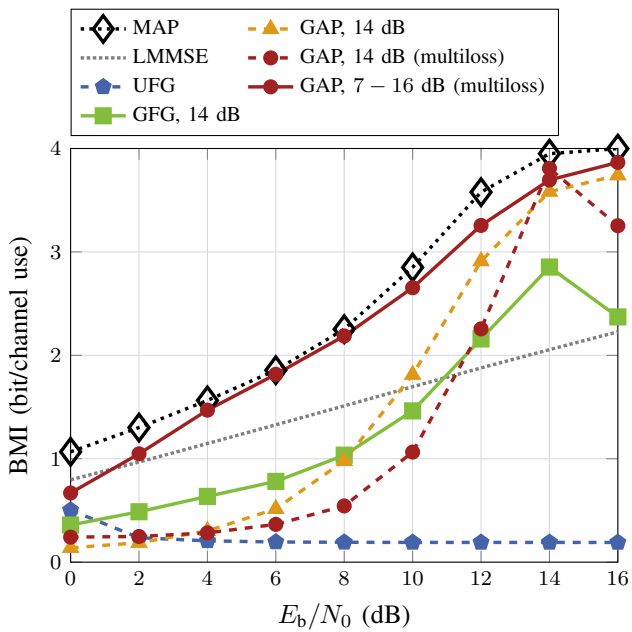
\begin{figure}[tb]
\centering
    \begin{tikzpicture}[trim axis right]
    \begin{axis}[
          width=\columnwidth,
          height=0.8\columnwidth,
          grid=major, %
          grid style={gray!30}, %
          xlabel= $E_\text{b}/N_0$ (dB),
          ylabel= BMI (bit/channel use),
          enlarge x limits=false,
          enlarge y limits=false,
          ymin = 0.0,
          ymax = 4.0,
          legend style={at={(0.0, 1.03)},anchor=south west, font=\footnotesize},
          legend cell align={left},
          legend columns = 4,
          transpose legend,
          tick label style={font=\footnotesize},
        ]
        \addplot[dotted, mark=diamond,mark options={scale=2.5,solid}, color=black, line width=1.3pt] table[x=Eb/N0, y=BCJR_BMI ,col sep=comma] {numerical_results/proakisB/16QAM.csv};
        \addplot[densely dotted, color=gray, line width=1.3pt] table[x=Eb/N0, y=MMSE_BMI ,col sep=comma] {numerical_results/proakisB/16QAM.csv};
        \addplot[dashed, mark=pentagon*,mark options={scale=1.3,solid}, color=KITblue, line width=1.3pt] table[x=Eb/N0, y=UB_FGEQ_BMI ,col sep=comma] {numerical_results/proakisB/16QAM.csv};
        \addplot[mark=square*, color=KITpalegreen, line width=1.3pt, mark options={scale=1.3,solid}] table[x=Eb/N0, y=GFG_BMI ,col sep=comma] {numerical_results/proakisB/16QAM.csv};
        \addplot[dashed, mark=triangle*,mark options={scale=1.3,solid}, color=KITorange, line width=1.3pt] table[x=Eb/N0, y=GFG2D_14dB_BMI ,col sep=comma] {numerical_results/proakisB/16QAM.csv};
        \addplot[dashed, mark=*,mark options={scale=1.1,solid}, color=KITred, line width=1.3pt] table[x=Eb/N0, y=GFG2D_14dB_multiloss_BMI ,col sep=comma] {numerical_results/proakisB/16QAM.csv};
        \addplot[mark=*,mark options={scale=1.1,solid}, color=KITred, line width=1.3pt] table[x=Eb/N0, y=GFG2D_7-16dB_multiloss_BMI ,col sep=comma] {numerical_results/proakisB/16QAM.csv};
        
        \legend{MAP \\ LMMSE \\ UFG \\ GFG, 14~dB  \\ GAP, 14~dB \\ GAP, 14~dB (multiloss)\\ GAP, $7-16$~dB (multiloss)\\ }
    \end{axis}
    \end{tikzpicture}
    \caption{BMI versus $\ebno$ of different symbol detectors for a 16-\ac{QAM} transmission over the channel $\text{C}_\text{B}$. The GAP algorithm is parametrized with ${(S,B,N') = (5,2,4)}$. The GFG and the GAP algorithm both apply preprocessors of length ${L_\text{p}=7}$.}
    \label{fig:proakisB_16QAM_BMI}
\end{figure}
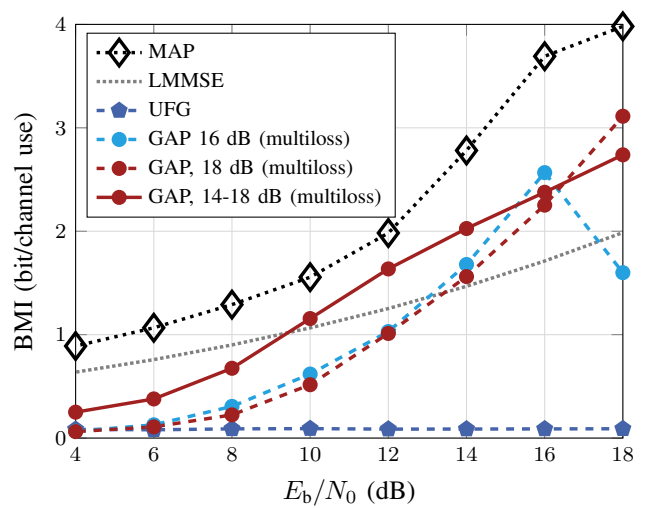
\begin{figure}[tb]
\centering
    \begin{tikzpicture}[trim axis right]
    \begin{axis}[
          width=\columnwidth,
          height=0.8\columnwidth,
          grid=major, %
          grid style={gray!30}, %
          xlabel= $E_\text{b}/N_0$ (dB),
          ylabel= BMI (bit/channel use),
          enlarge x limits=false,
          xtick = {4,6,8,10,12,14,16,18},
          enlarge y limits=false,
          xmin = 4,
          ymin = 0.0,
          ymax = 4.0,
          legend style={at={(0.02, 0.98)},anchor=north west, font=\footnotesize},
          legend cell align={left},
          tick label style={font=\footnotesize},
        ]
        \addplot[dotted, mark=diamond,mark options={scale=2.5,solid}, color=black, line width=1.3pt] table[x=Eb/N0, y=BCJR_BMI ,col sep=comma] {numerical_results/proakisC/16QAM.csv};
        \addplot[densely dotted, color=gray, line width=1.3pt] table[x=Eb/N0, y=MMSE_BMI ,col sep=comma] {numerical_results/proakisC/16QAM.csv};
        \addplot[dashed, mark=pentagon*,mark options={scale=1.3,solid}, color=KITblue, line width=1.3pt] table[x=Eb/N0, y=UB_FGEQ_BMI ,col sep=comma] {numerical_results/proakisC/16QAM.csv};
        \addplot[dashed, mark=*,mark options={scale=1.1,solid}, color=KITcyan, line width=1.3pt] table[x=Eb/N0, y=GFG2D_16dB_multiloss_BMI ,col sep=comma] {numerical_results/proakisC/16QAM.csv};
        \addplot[dashed, mark=*,mark options={scale=1.1,solid}, color=KITred, line width=1.3pt] table[x=Eb/N0, y=GFG2D_18dBv2_multiloss_BMI ,col sep=comma] {numerical_results/proakisC/16QAM.csv};
        
        \addplot[mark=*,mark options={scale=1.1,solid}, color=KITred, line width=1.3pt] table[x=Eb/N0, y=GFG2D_14-18dB_multiloss_BMI ,col sep=comma] {numerical_results/proakisC/16QAM.csv};
        
        \legend{MAP \\ LMMSE \\ UFG\\ GAP 16~dB (multiloss) \\ GAP, 18~dB (multiloss) \\ GAP, 14-18~dB (multiloss) \\ }
    \end{axis}
    \end{tikzpicture}
    \caption{BMI over $\ebno$ of the GAP algorithm with ${(S,B,N') = (5,2,4)}$ and ${L_\text{p}=9}$ for a 16-\ac{QAM} transmission over the channel $\text{C}_\text{C}$, optimized for different ranges of $\ebno$.}
    \label{fig:proakisC_16QAM_BMI}
\end{figure}

Finally, we evaluate the GAP algorithm with ${(S,B,N')=(5,2,4)}$ for a 16-\ac{QAM} constellation over the channels $\text{C}_\text{B}$ and $\text{C}_\text{C}$.
The simulation results for the channel $\text{C}_\text{B}$ are reported in Fig.~\ref{fig:proakisB_16QAM_BMI}. Relevant for practical applications of a 16-\ac{QAM} signaling is the $\ebno$ range around ${10-16}$~dB, where the \ac{MAP} detector approaches the upper \ac{BMI} bound of $4$~bit/channel~use. Optimizing $\mathcal{P}_\text{GAP}$ with ${L_\text{p}=7}$ for ${\ebno = 14}$~dB in an end-to-end manner yields a \ac{BMI} of $3.58$~bit/channel~use.
The performance can be further improved to a BMI of $3.81$~bit/channel~use for ${\ebno = 14}$~dB by employing the multiloss term \eqref{eq:multiloss} for the optimization process.
The GAP detector outperforms the GFG algorithm by $0.95$~bit/channel~use and the \ac{MMSE} equalizer by $1.75$~bit/channel~use, thereby closing the gap to optimum performance.
However, both algorithms do not generalize well, especially for the low $\ebno$ regime. 
To reduce overfitting to a specific $\ebno$, we can vary the $\ebno$ while optimizing $\mathcal{P}_\text{GAP}$. For instance, sampling the $\ebno$ from a uniform distribution in the range ${7-16}$~dB during training consequently results in an improved average performance of the GAP algorithm in this $\ebno$ range. Additionally, the detection performance also significantly improves for $\ebno < 7$~dB, even though this $\ebno$ range was not sampled during the optimization. The generalized training leads to a minor BMI degradation of $0.11$~bit/channel~use for ${\ebno = 14}$~dB, compared to the optimization at a fixed ${\ebno = 14}$~dB. The training range of the $\ebno$ should thus match the region of operation of the detector as accurately as possible.
Further increasing the number of stages $S$, branches $B$, iterations $N'$ or the filter order $L_\text{p}$ has not shown any significant performance gain.
The BMI performance of the GAP algorithm for the channel $C_\text{C}$ and 16-\ac{QAM} is reported in Fig.~\ref{fig:proakisC_16QAM_BMI}. 
We evaluate three different GAP instances, which were all optimized w.r.t. the \ac{BMI} multiloss term \eqref{eq:multiloss}, but differ in the range from which the $\ebno$ was uniformly sampled during the training. The BMI performance of the GAP detector behaves qualitatively similar to the results on channel $C_\text{B}$. Trained for a specific $\ebno$, e.g., ${\ebno=18}$~dB, the GAP algorithm outperforms the \ac{MMSE} equalizer about $1.1$~bit/channel~use in terms of the \ac{BMI}.

\section{Conclusion}\label{sec:conclusion}
We studied the application and neural enhancement of factor graph-based symbol detectors on \ac{AWGN} channels with linear \ac{ISI}.
We proposed simple but effective generalizations of the factor graph, 
as well as \ac{NBP} as an enhanced message passing algorithm in order to mitigate the effect of cycles in the graphs.
The methods are only marginally increasing the detection complexity compared to the UFG algorithm. 
We further proposed the novel symbol detection algorithm GAP which comprises both \ac{NBP} as well
as a dynamic transition of the underlying factor graph.
The algorithm delivers an attractive and highly scalable tradeoff between performance and complexity.
Our methods showed a significant performance improvement of the factor graph-based symbol detector, closing the gap to optimum detection performance in  various transmission scenarios.
Especially for high-order constellations and static channels with large memory, the proposed GAP algorithm is a promising low-complexity alternative to the BCJR algorithm.

\end{document}